\documentclass{jfm}
\usepackage{graphicx}
\usepackage[font=small,skip=0pt]{caption}
\usepackage{titlesec}
\titlespacing*{\section}{0pt}{1.1\baselineskip}{\baselineskip}
\usepackage{epstopdf, epsfig}
\usepackage[euler]{textgreek}
\usepackage{fixltx2e}
\usepackage{todonotes}
\usepackage{caption}
\usepackage{subfig}
\usepackage{tabularx}
\usepackage{wrapfig}
\usepackage{comment}
\usepackage{amsmath} 
\usepackage{cleveref}
\usepackage{amssymb}
\usepackage{amsfonts}
\usepackage{mathabx}
\usepackage{MnSymbol}
\usepackage{wasysym}

\usepackage{hhline}
\usepackage{multirow}
\usepackage{physics}
\usepackage{tabularx}
\usepackage{array}
\usepackage{subfig}
\usepackage{mathtools}
\usepackage{siunitx}
\newcolumntype{M}[1]{>{\centering\arraybackslash}m{#1}}
\usepackage{pdfpages}
\usepackage[running]{lineno}
\usepackage{float}

\newcommand{\ddtt}[1]{\frac{\text{d}#1(t)}{\text{d}t}}

\title{Neural Network aided quarantine control model estimation of global Covid-19 spread}
\author{ Raj Dandekar\aff{1} \and George Barbastathis\aff{2,3} \corresp{\email{gbarb@mit.edu}}}
\affiliation{\aff{1}Department of Civil and Environmental Engineering, Massachusetts Institute of Technology, Cambridge, MA 02139, USA, \aff{2} Department of Mechanical Engineering, Massachusetts Institute of Technology, Cambridge, MA 02139, USA, \aff{3}Singapore-MIT Alliance for Research and Technology (SMART) Centre, Singapore 138602} 

\begin{document}
\maketitle
Since the first recording of what we now call Covid-19 infection in Wuhan, Hubei province, China on Dec 31, 2019 \citep{casezero}, the disease has spread worldwide and met with a wide variety of social distancing and quarantine policies. The effectiveness of these responses is notoriously difficult to quantify as individuals travel, violate policies deliberately or inadvertently, and infect others without themselves being detected \citep{li2020early, wu2020nowcasting, wang2020phase, chinazzi2020effect, ferguson2020impact, kraemer2020effect}. Moreover, the publicly available data on infection rates are themselves unreliable due to limited testing and even possibly under-reporting \citep{li2020substantial}. In this paper, we attempt to interpret and extrapolate from publicly available data using a mixed first-principles epidemiological equations and data-driven neural network model. Leveraging our neural network augmented model, we focus our analysis on four locales: Wuhan, Italy, South Korea and the United States of America, and compare the role played by the quarantine and isolation measures in each of these countries in controlling the effective reproduction number $R_{t}$ of the virus. Our results unequivocally indicate that the countries in which rapid government interventions and strict public health measures for quarantine and isolation were implemented were successful in halting the spread of infection and prevent it from exploding exponentially. In the case of Wuhan especially, where the available data were earliest available, we have been able to test the predicting ability of our model by training it from data in the January $24^{\text{th}}$ till March $3^{\text{rd}}$ window, and then matching the predictions up to April $1^{\text{st}}$. Even for Italy and South Korea, we have a buffer window of one week ($25$ March - $1$ April) to validate the predictions of our model. In the case of the US, our model captures well the current infected curve growth and predicts a halting of infection spread by $20$ April 2020. We further demonstrate that relaxing or reversing quarantine measures right now will lead to an exponential explosion in the infected case count, thus nullifying the role played by all measures implemented in the US since mid March 2020.

\section{Introduction}
The Coronavirus respiratory disease 2019 originating from the virus ``SARS-CoV-2" \citep{chan2020familial, cdc} has led to a global pandemic, leading to $823,626$ confirmed global cases in more than $200$ countries as of April 1, 2020 \citep{worldcoronavirus}. 
As the disease began to spread beyond its apparent origin in Wuhan, the responses of local and national governments varied considerably. The evolution of infections has been similarly diverse, in some cases appearing to be contained and in others reaching catastrophic proportions. In Hubei province itself, starting at the end of January, more than $10$~million residents were quarantined by shutting down public transport systems, train and airport stations, and imposing police controls on pedestrian traffic. Subsequently, similar policies were applied nation-wide in China. By the end of March, the rate of infections was reportedly receding \citep{cyranoski2020china}. Taiwan, Hong Kong, and Singapore managed to maintain fairly low infection rates throughout, even though a second wave of infections is appearing in Singapore, perhaps due to incoming repatriates \citep{singapore}. South Korea, Iran, Italy, and Spain experienced acute initial increases, but then adopted drastic generalized quarantine. This did result in apparent recession of the spread in South Korea, whereas in the other three countries the effect of the policies is not yet clear. In the United States, where both the onset of widespread infections and government responses were comparatively delayed, infection growth currently appears to be explosive. As of April $2$ 2020, the United States has the highest number of infected cases ($\sim 227\text{k}$) globally. Given the available data by country and world-wide, there is an urgent need to use data driven approaches to quantitatively estimate and compare the role of the quarantine policy measures implemented in several countries in curtailing spread of the disease. 

Existing models analyzing the role of travel restrictions in the spread of Covid-19 either used parameters based on prior knowledge of SARS/MERS coronavirus epidemiology and not derived independently from the Covid-19 data \citep{chinazzi2020effect}, or were not implemented on a global scale \citep{kraemer2020effect}. In this paper, we propose augmenting a first principles-derived epidemiological model with a data-driven module, implemented as a neural network. We leverage this model to analyze and compare the role of quarantine control policies employed in Wuhan, Italy, South Korea and USA, in controlling the virus effective reproduction number $R_{t}$ \citep{imai2020report, read2020novel, tang2020estimation, li2020early, wu2020nowcasting, kucharski2020early, ferguson2020impact}. In the original model, known as SEIR \citep{SEIR1, SEIR2, SEIR3}, the population is divided into the susceptible $S$, exposed $E$, infected $I$ and recovered $R$ groups, and their relative growths and competition are represented as a set of coupled ordinary differential equations. The simpler SIR model does not account for the exposed population $E$. These models cannot capture the large-scale effects of more granular interactions, such as the population's response to social distancing and quarantine policies. This is where data come in: in our approach, a neural network added as a non linear function approximator \citep{Rackauckas20} informs the infected variable $I$  in the SIR model. This neural network encodes information about the quarantine strength function in the locale where the model is implemented. The neural network is trained from publicly available infection and population data for Covid-19 for a specific region under study; details are in the Materials and Methods section. Thus, our proposed model is globally applicable and interpretable with parameters learnt from the current Covid-19 data, and does not rely upon data from previous epidemics like SARS/MERS.

Since neural networks can be used to approximate nonlinear functions with a finite set of parameters, they serve as a powerful tool to approximate quarantine effects in combination with the analytical epidemiological models. The downside is that the internal workings of a neural network are difficult to interpret. The recently emerging field of Physics-Informed Neural Networks \citep{raissi2019physics} exploits conservation principles, SIR in our case, to mitigate overfitting and other related machine learning risks. 

All four regions that we applied our model to have developed infected and exposed populations that are sufficiently large to train our models. The first three are comparable in terms of population ($11$ million, $60$ million and $52$ million, respectively) and almost complete isolation from inbound travel while the USA has a much larger population ($327$ million), with increasing travel restrictions since mid March 2020. Leveraging the insights gained through reliable prediction and estimation in Wuhan, South Korea and Italy, we make forecasting predictions regarding the infection spread in the USA; thus making our model informative for quarantine and social distancing policy guidelines and regulations.

\begin{figure}
\centering
\begin{tabular}{cc}
\subfloat[]{\includegraphics[width=0.4\textwidth]{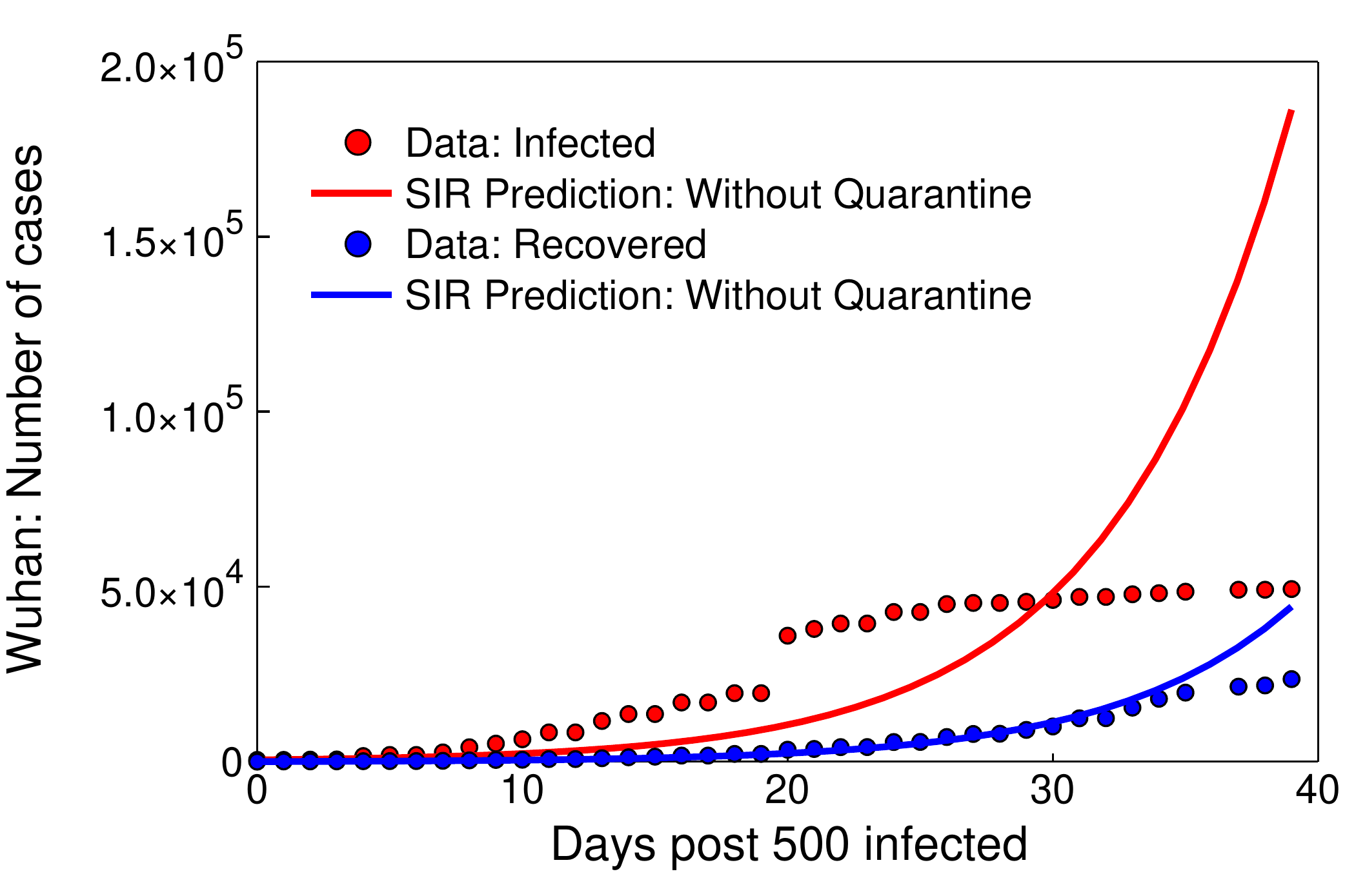}}
\subfloat[]{\includegraphics[width=0.4\textwidth]{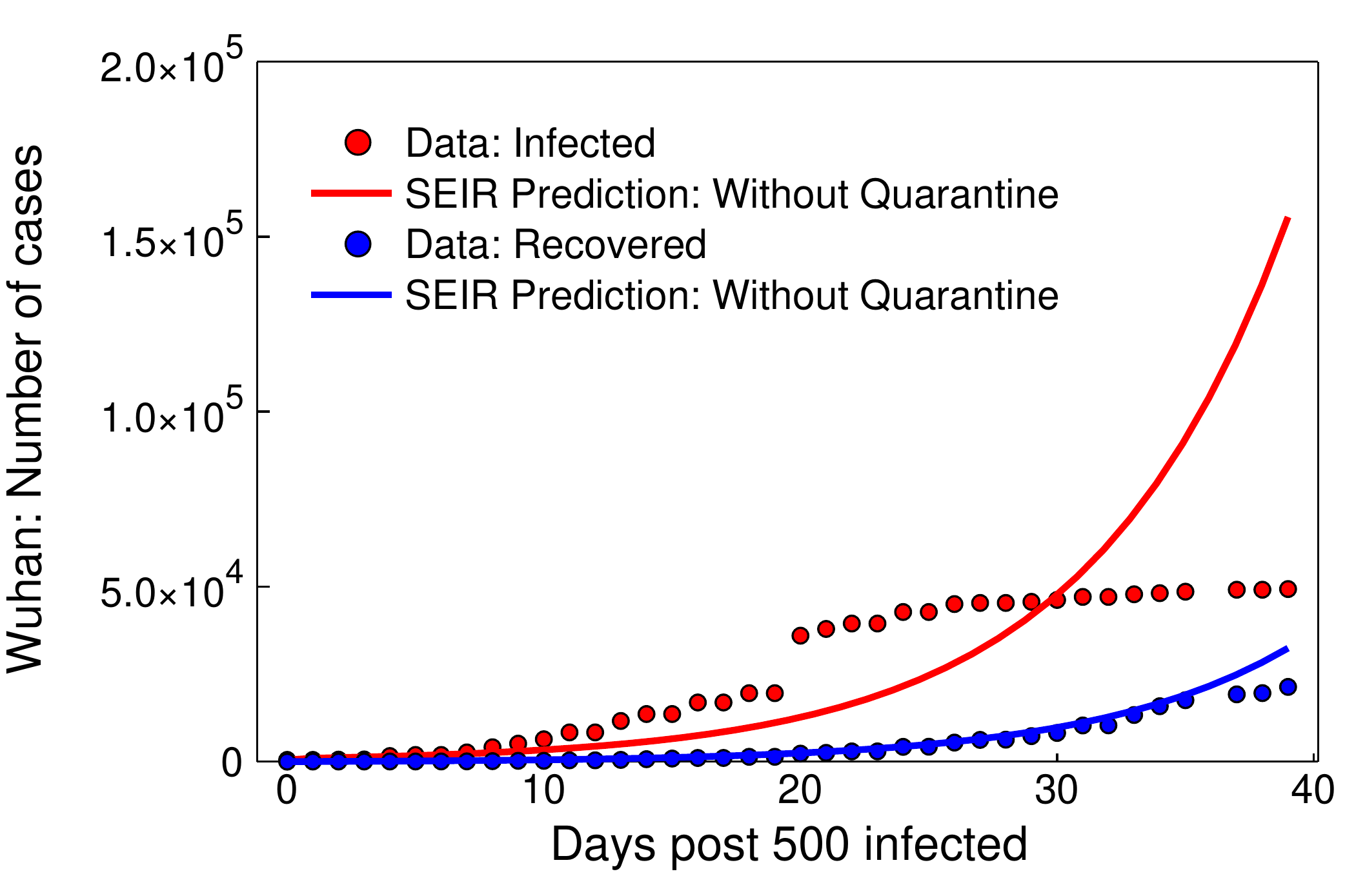}}\\
\end{tabular}
\caption{\textbf{Wuhan without quarantine control; day 0 = 24 January 2020:} Estimation of the infected and recovered case count compared to the data acquired from the Chinese National Health Commission based on the (a) SIR epidemiological model and (b) SEIR epidemiological model.}\label{figure1}
\end{figure}\label{Wuhan-no-quar}

\begin{figure}
\centering
\begin{tabular}{ccc}
\subfloat[]{\includegraphics[width=0.337\textwidth]{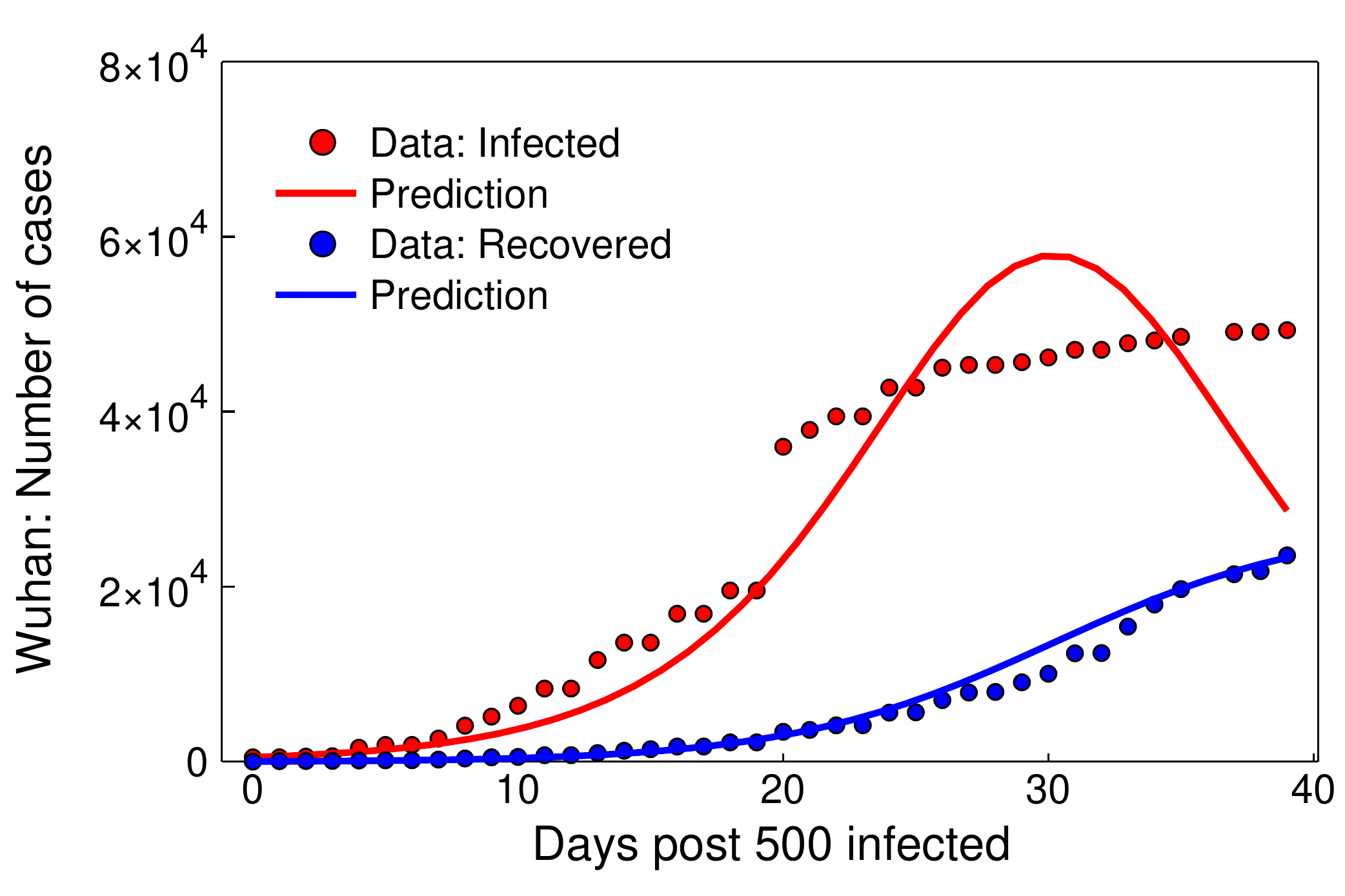}}
\subfloat[]{\includegraphics[width=0.32\textwidth]{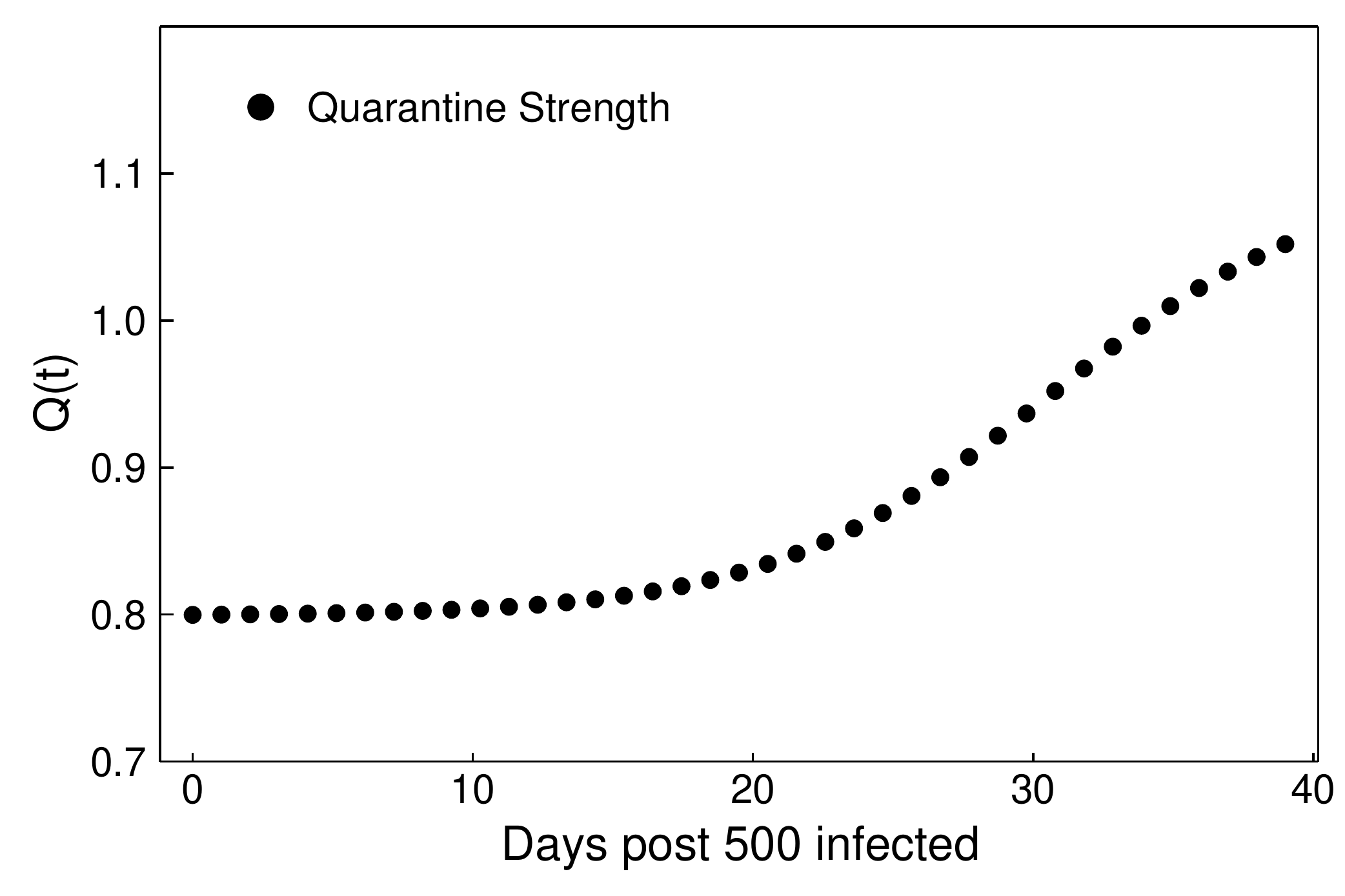}}
\subfloat[]{\includegraphics[width=0.32\textwidth]{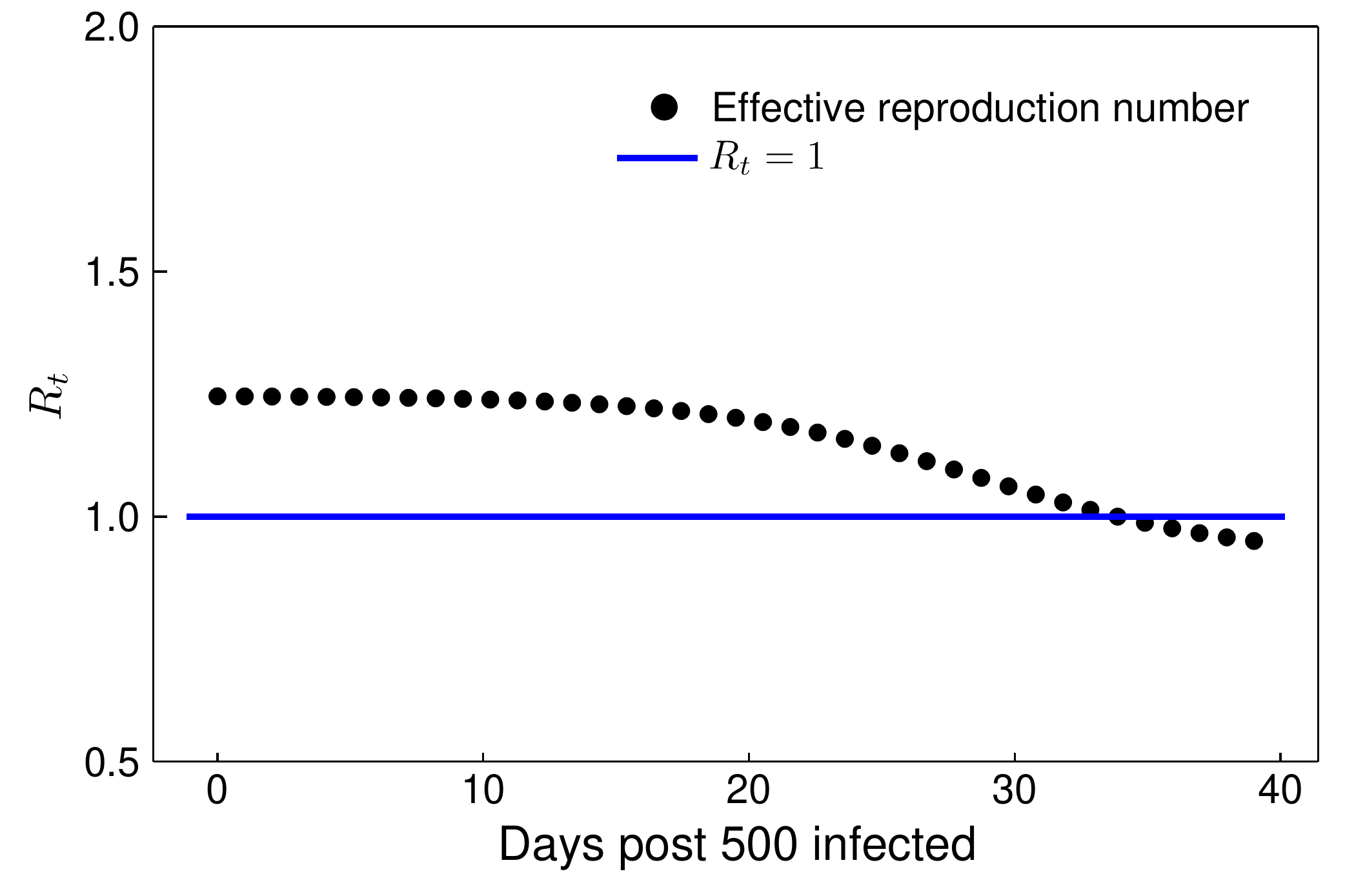}}
\end{tabular}
\caption{\textbf{Wuhan neural network model, day 0 = 24 January 2020:} (a) Estimation of the infected and recovered case count compared to the data. (b) Quarantine strength function $Q(t)$ learnt by the neural network. (c) Effective reproduction number $R_{t}$. }\label{Wuhan-quar}
\end{figure}

\begin{figure}
\centering
\begin{tabular}{cc}
\subfloat[]{\includegraphics[width=0.4\textwidth]{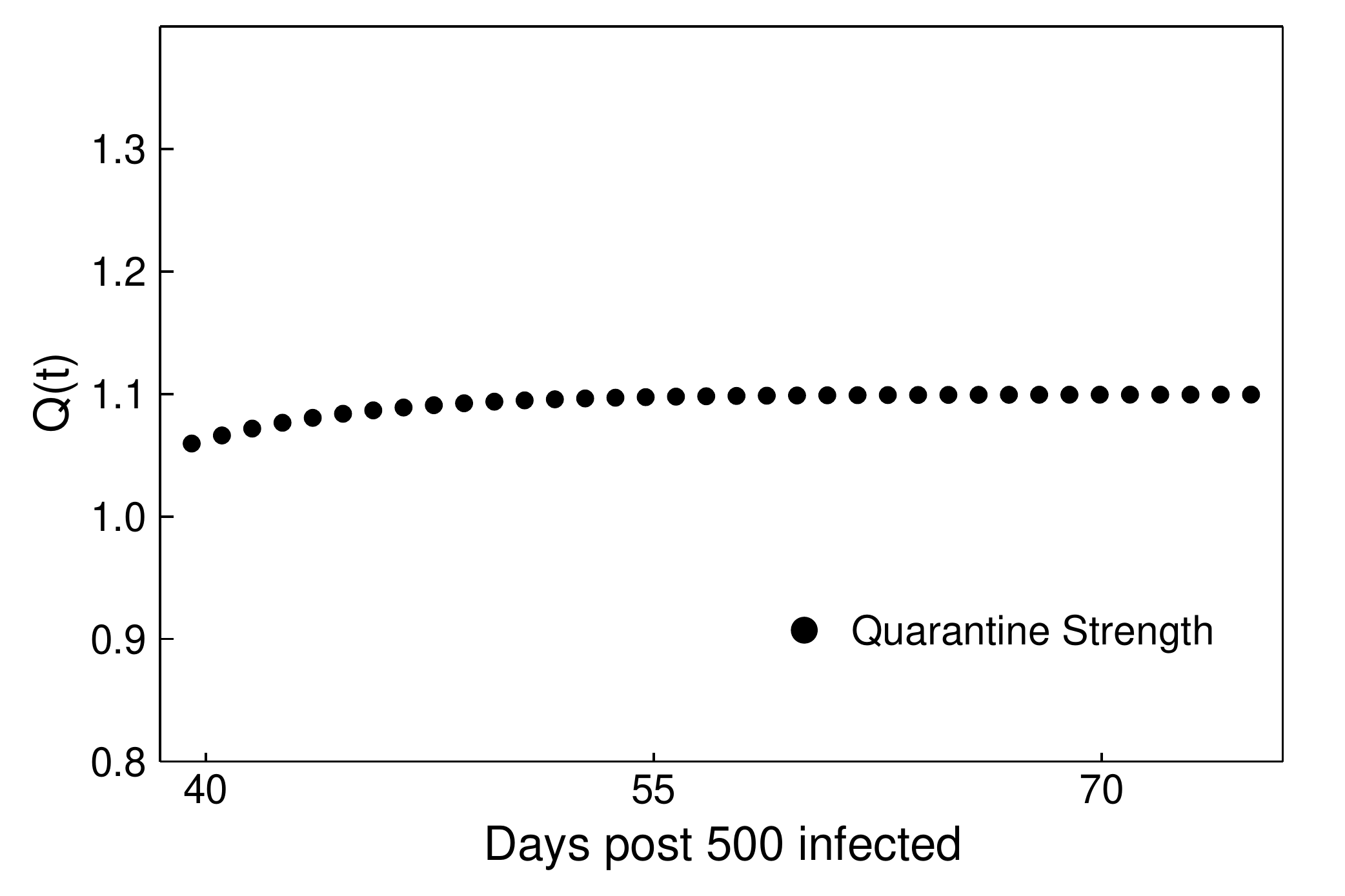}}
\subfloat[]{\includegraphics[width=0.4\textwidth]{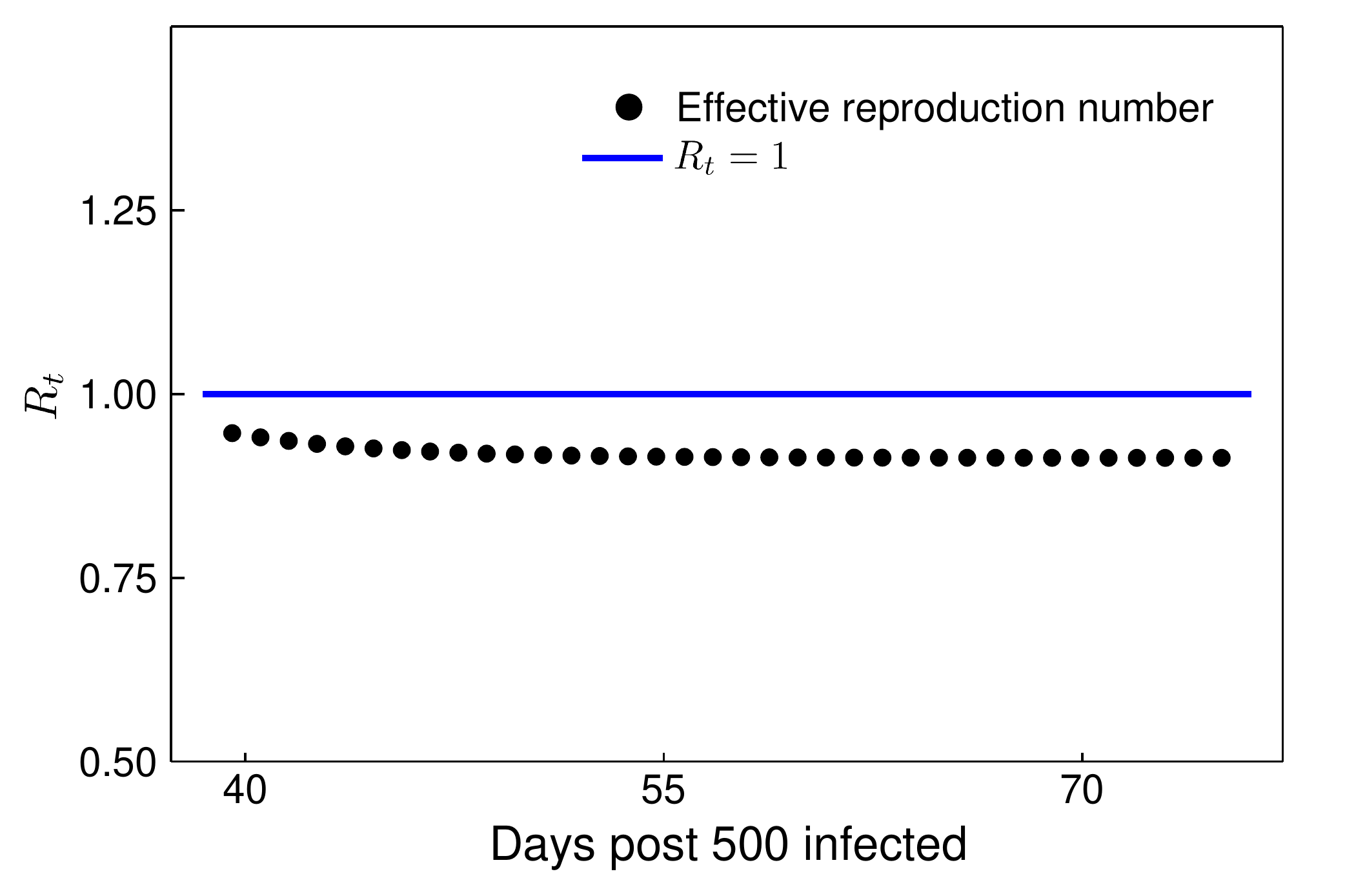}}
\end{tabular}
\caption{ \textbf{Wuhan forecasting:} (a) Quarantine strength $Q(t)$ and (b) Effective reproduction number $R_{t}$ based on the neural network augmented SIR model. }\label{Wuhan-1month-forecast}
\end{figure}

\begin{figure}
\centering
\begin{tabular}{ccc}
\subfloat[]{\includegraphics[width=0.33\textwidth]{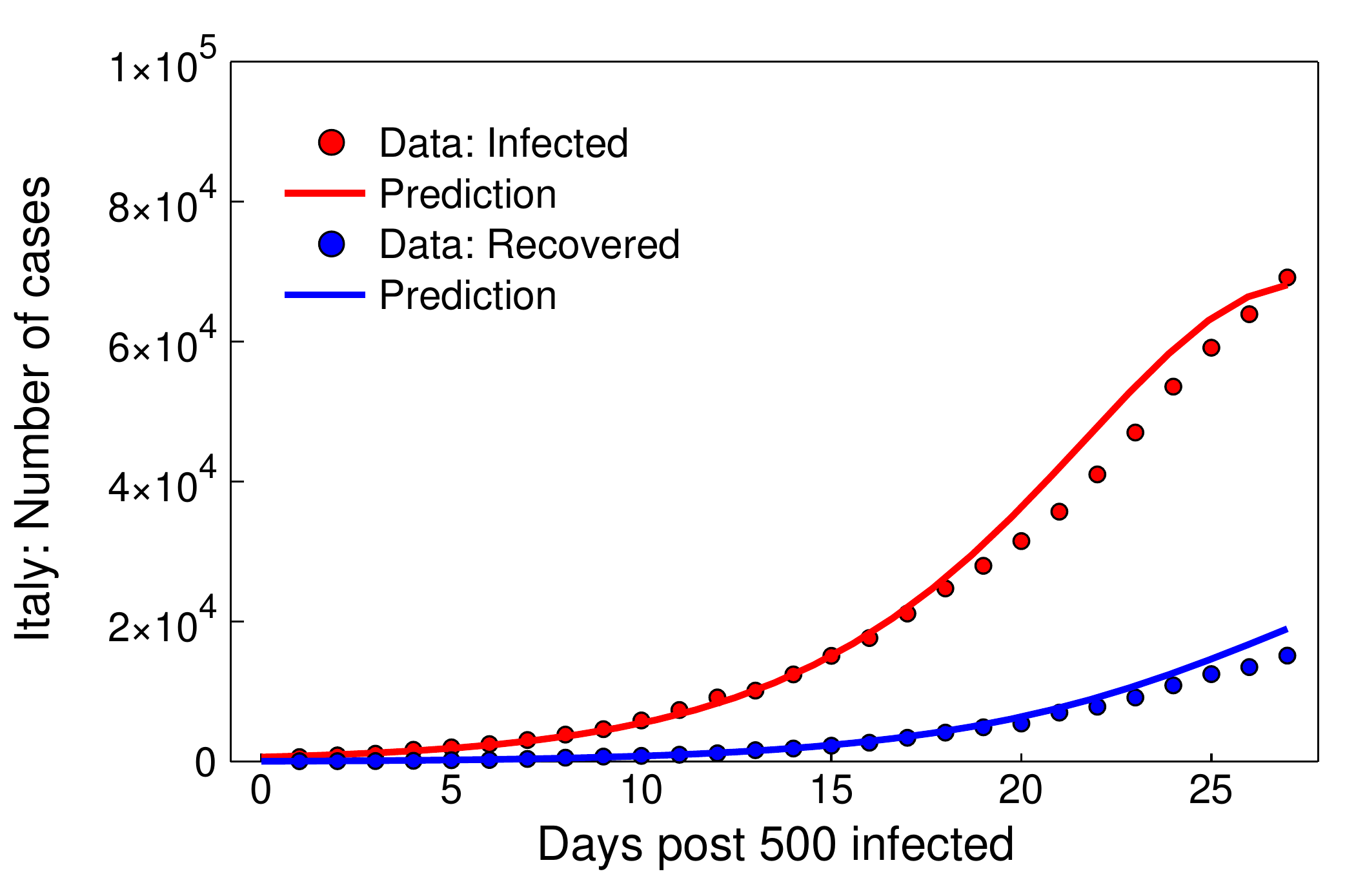}}
\subfloat[]{\includegraphics[width=0.315\textwidth]{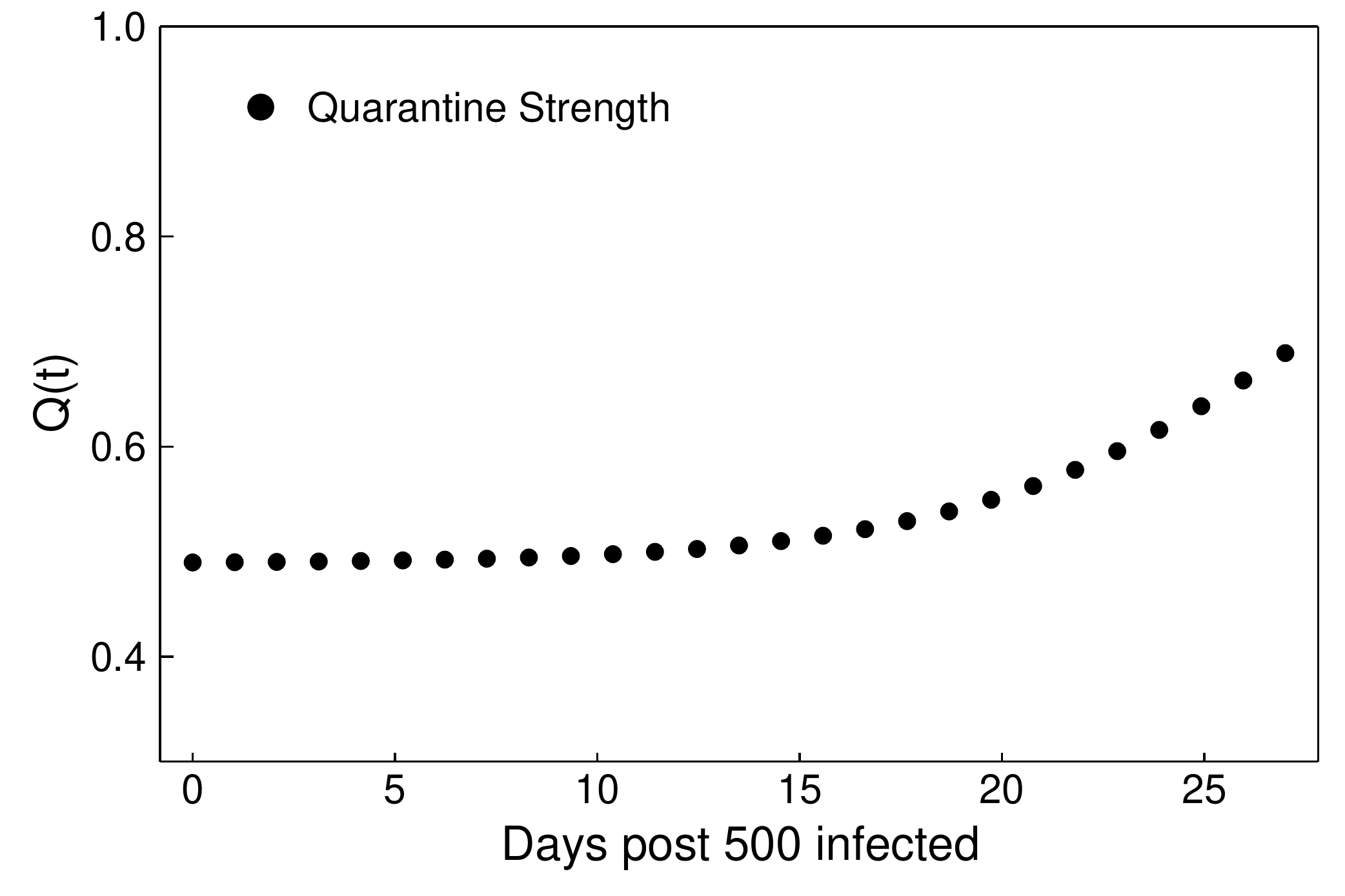}}
\subfloat[]{\includegraphics[width=0.315\textwidth]{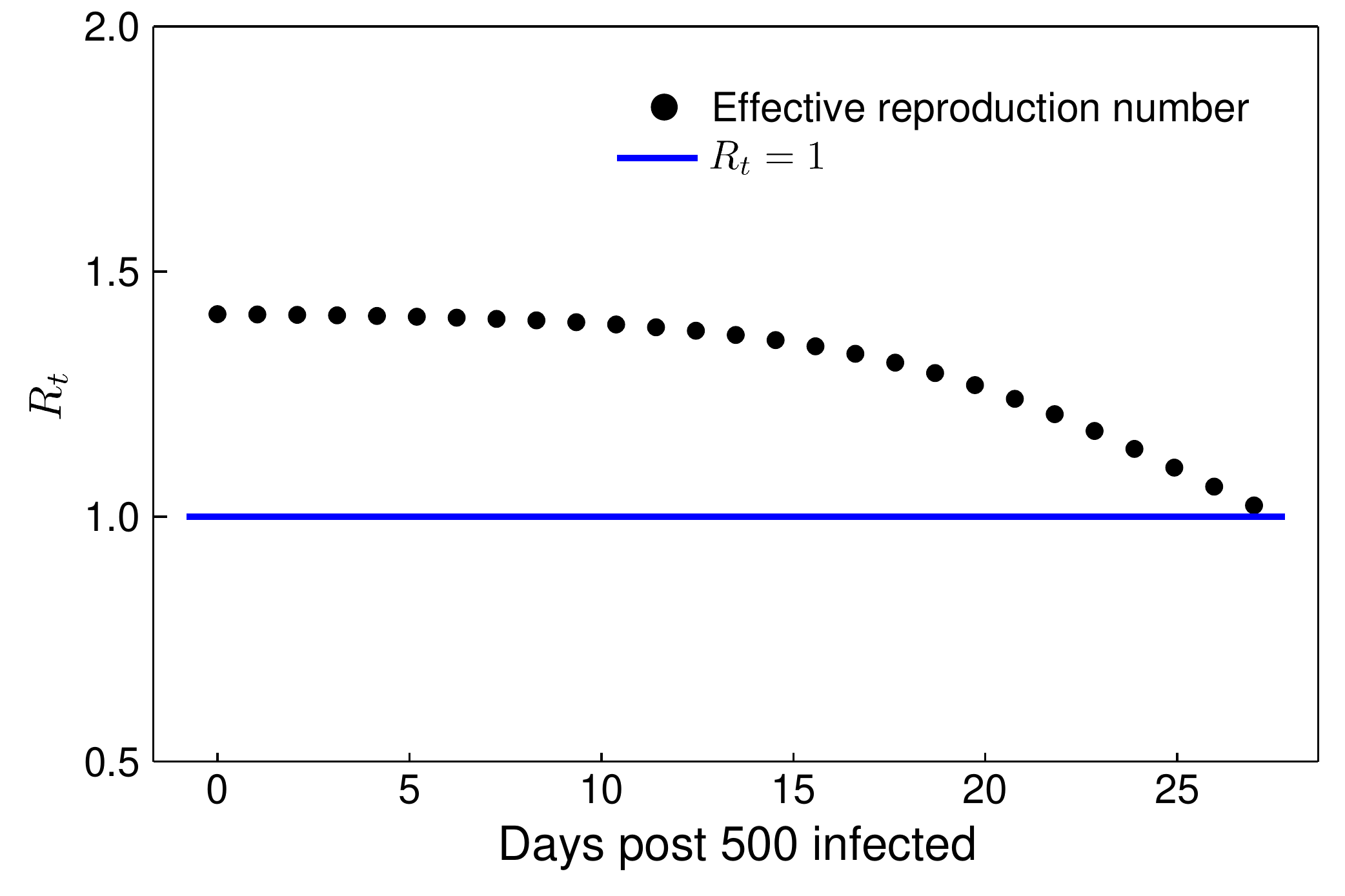}}
\end{tabular}
\caption{\textbf{Italy neural network model, day 0 = 27 February 2020:} (a) Estimation of the infected and recovered case count compared to the data. (b) Quarantine strength function $Q(t)$ learnt by the neural network. (c) Effective reproduction number $R_{t}$.}\label{Italy}
\end{figure}

\begin{figure}
\centering
\begin{tabular}{cc}
\subfloat[]{\includegraphics[width=0.4\textwidth]{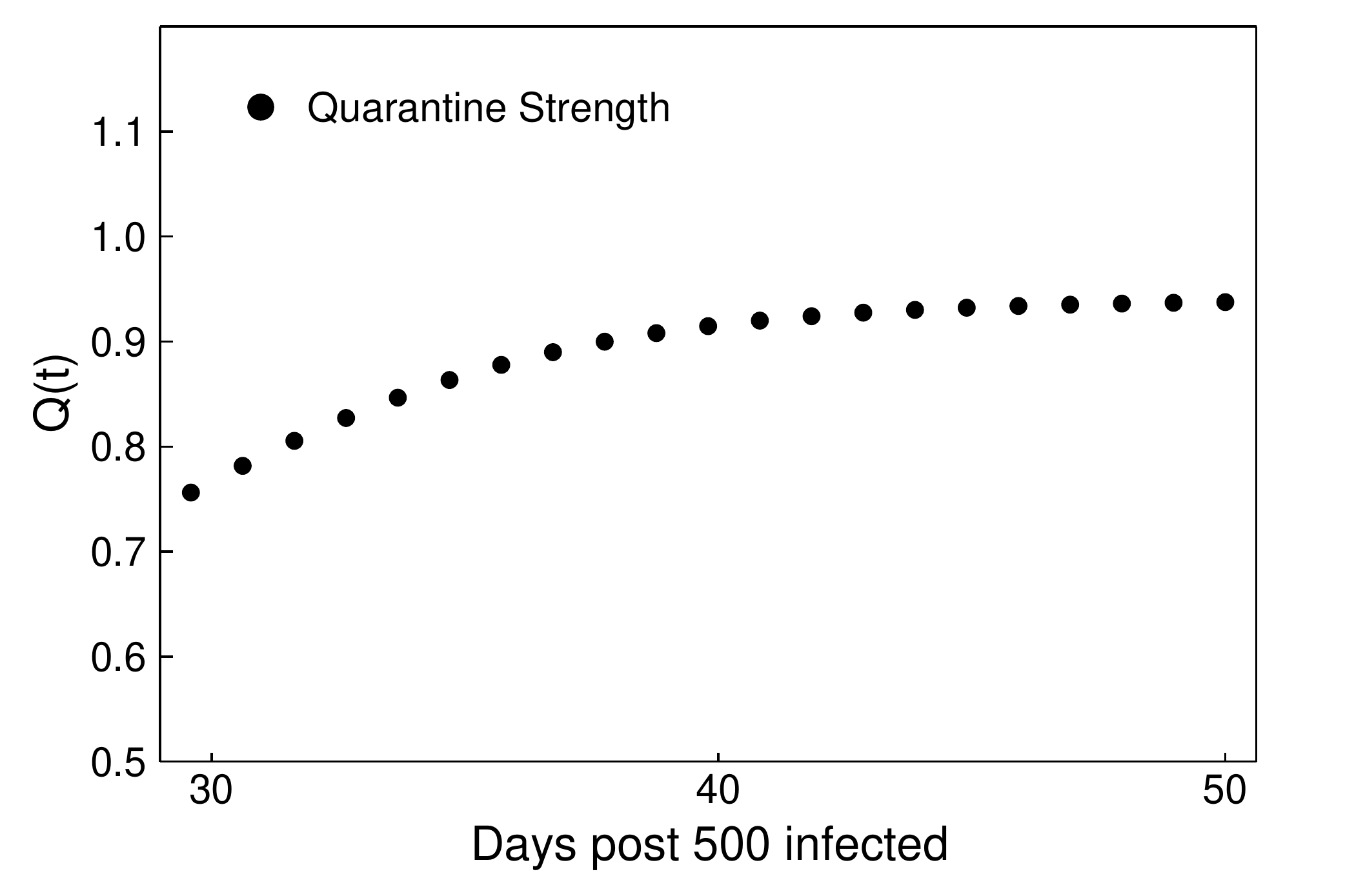}}
\subfloat[]{\includegraphics[width=0.4\textwidth]{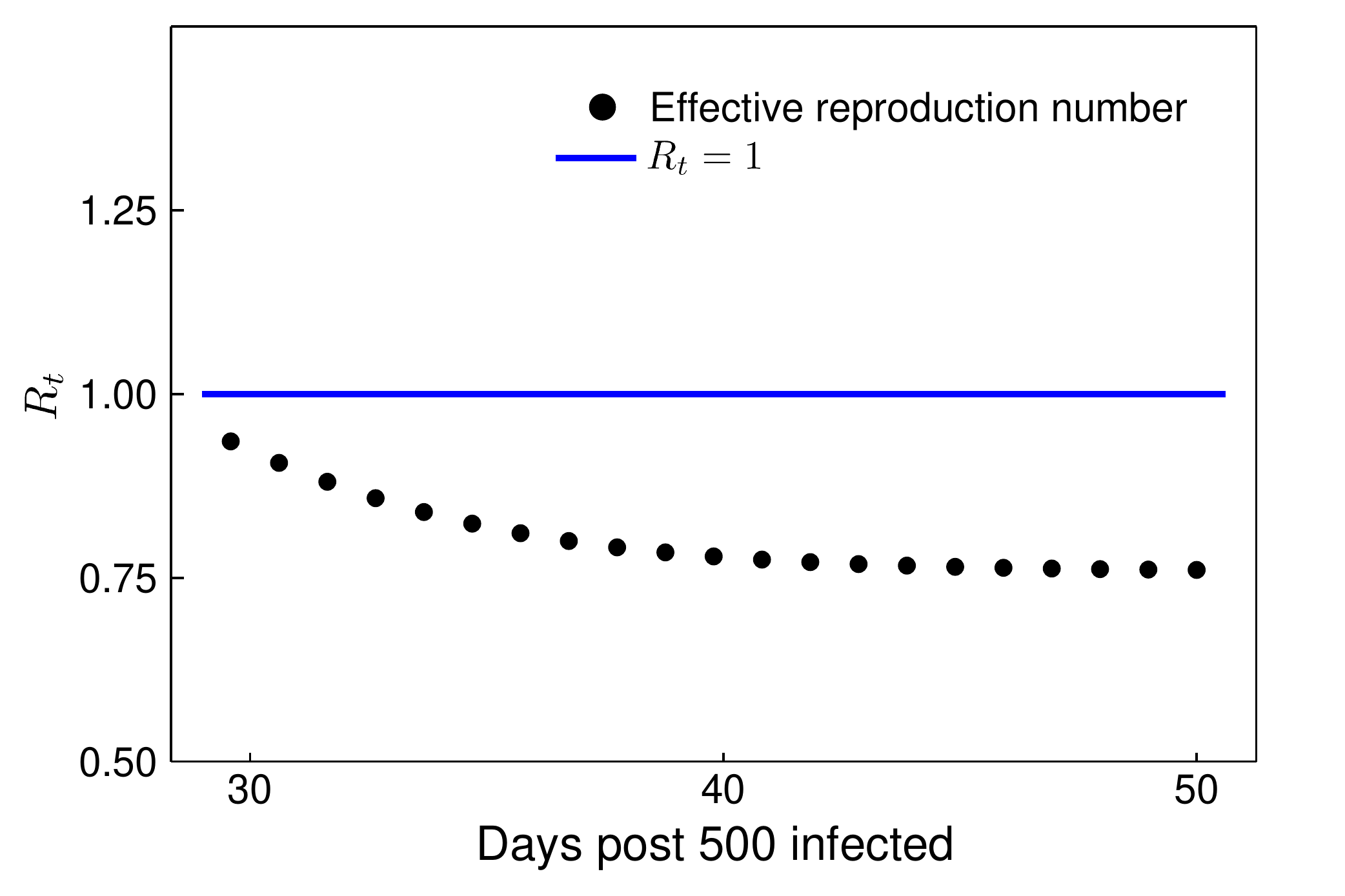}}
\end{tabular}
\caption{ \textbf{Italy forecasting:} (a) Quarantine strength, $Q$ and (b) Effective reproduction number, $R_{t}$ in Italy based on the neural network augmented SIR model. }\label{Italyf}
\end{figure}

\begin{figure}
\centering
\begin{tabular}{ccc}
\subfloat[]{\includegraphics[width=0.333\textwidth]{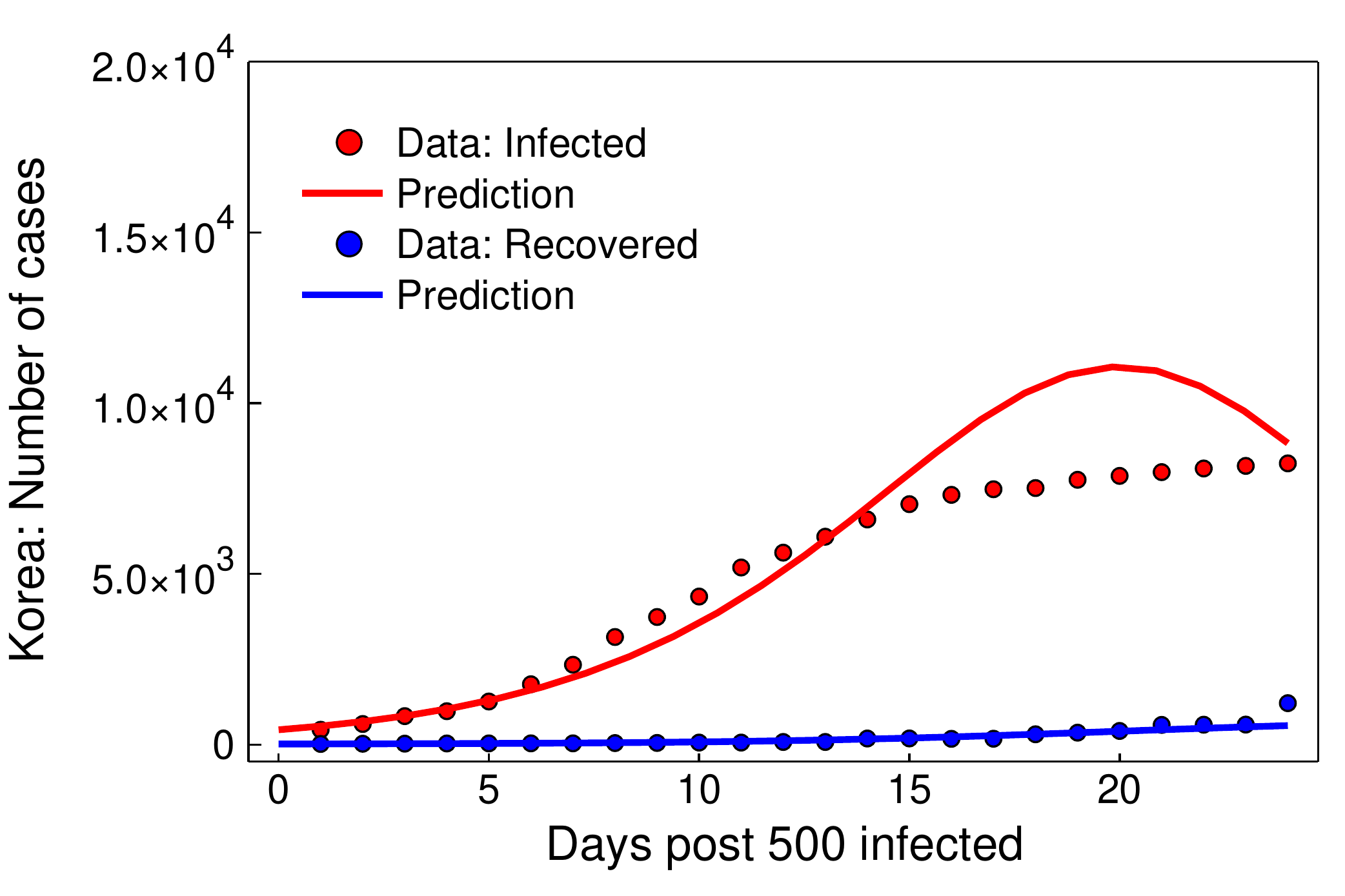}}
\subfloat[]{\includegraphics[width=0.32\textwidth]{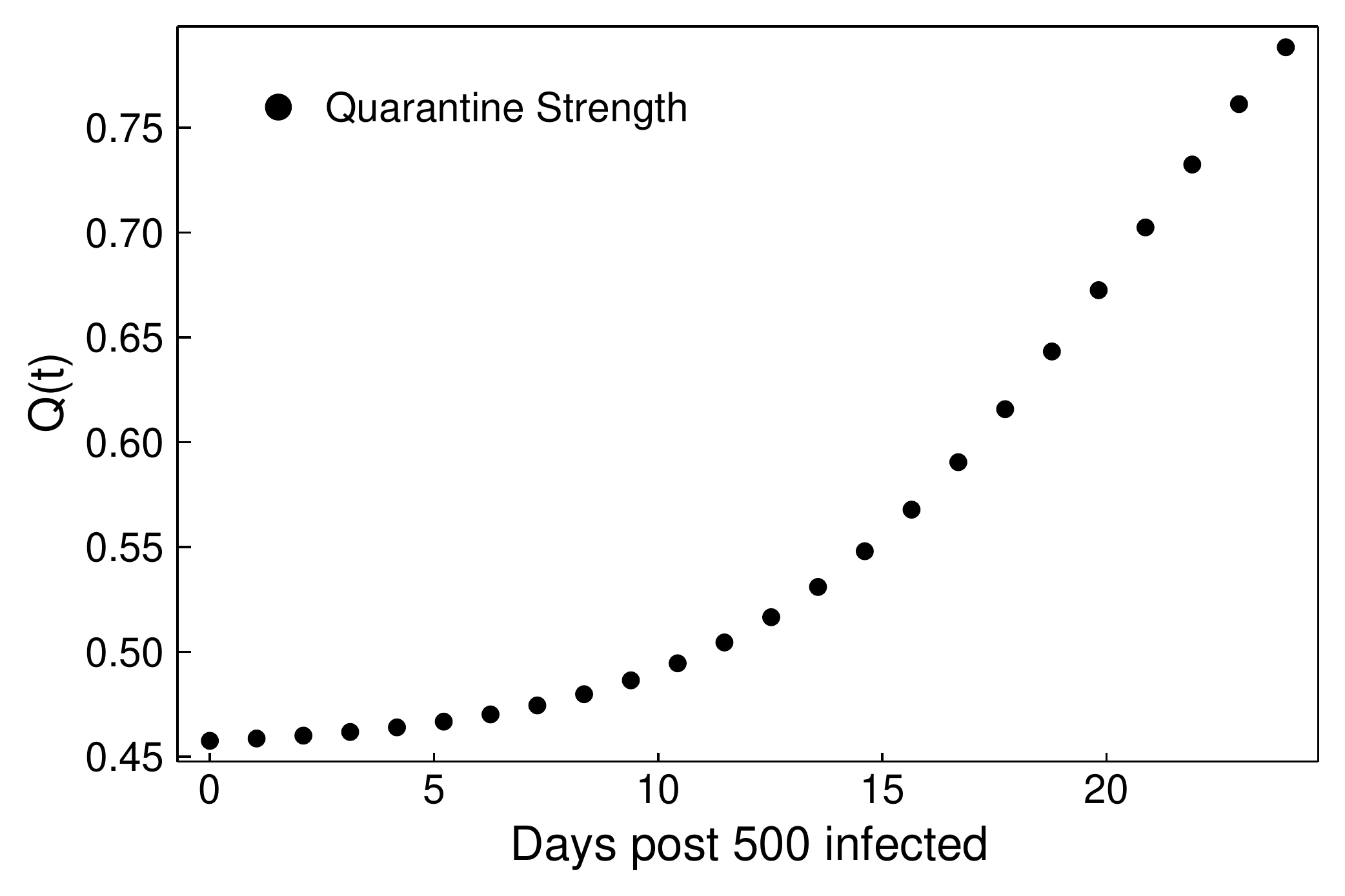}}
\subfloat[]{\includegraphics[width=0.32\textwidth]{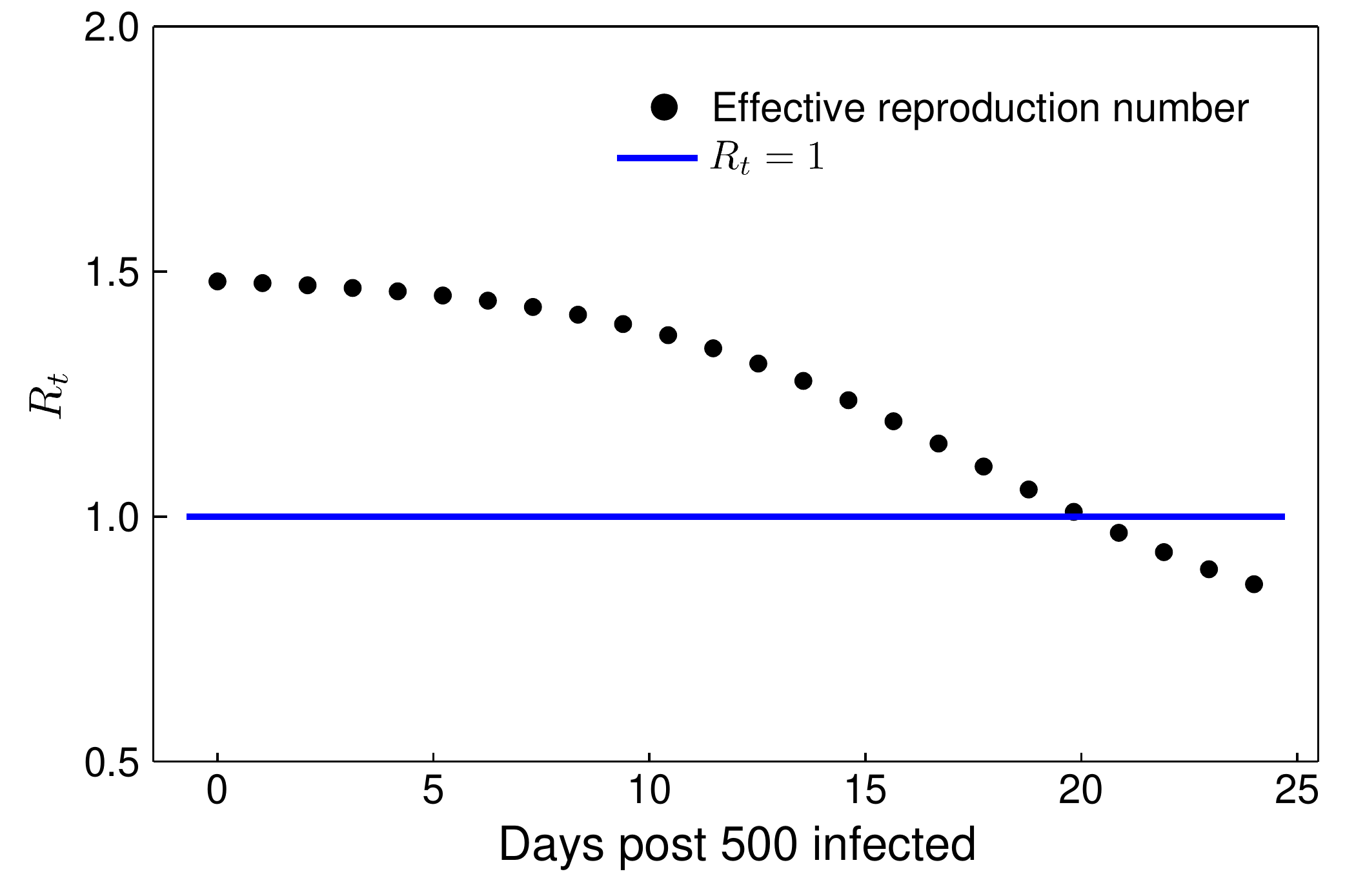}}
\end{tabular}
\caption{\textbf{South Korea neural network model, day 0 = 22 February 2020:} (a) Estimation of the infected and recovered case count compared to the data post 22 February 2020 in South Korea. (b) Quarantine strength function $Q(t)$ learnt by the neural network. (c) Effective reproduction number $R_{t}$.}\label{Korea}
\end{figure}

\begin{figure}
\centering
\begin{tabular}{cc}
\subfloat[]{\includegraphics[width=0.4\textwidth]{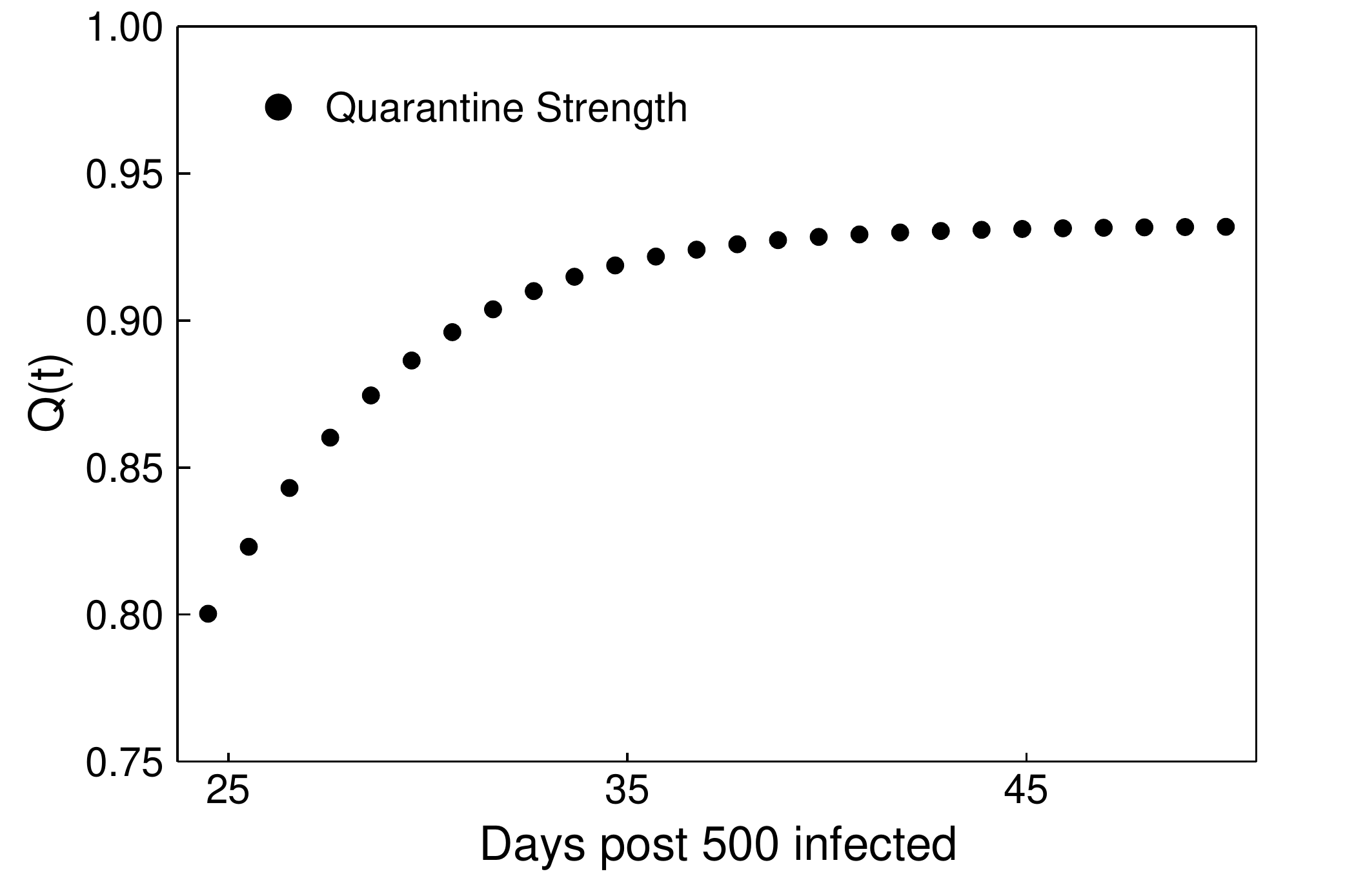}}
\subfloat[]{\includegraphics[width=0.4\textwidth]{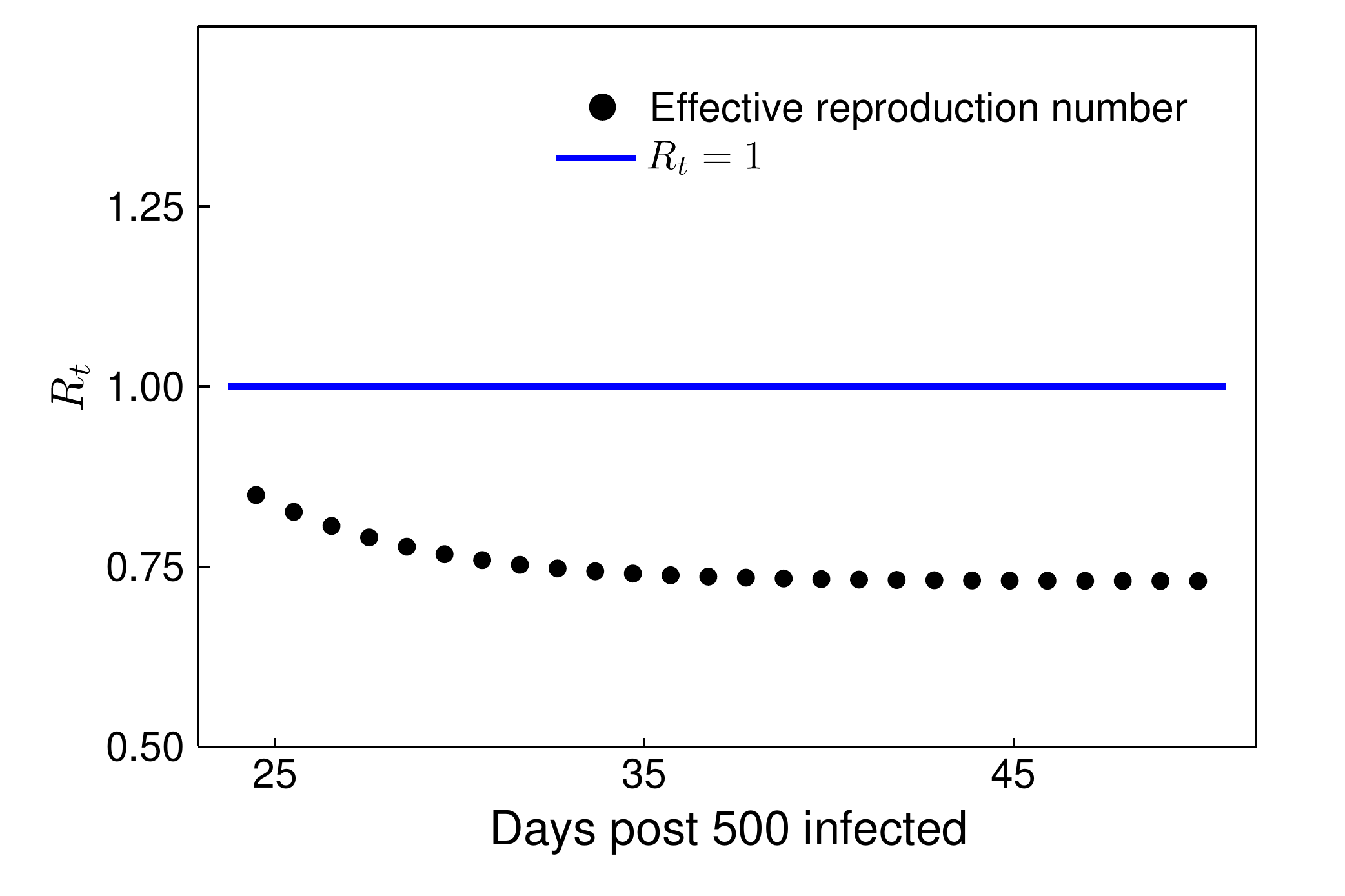}}
\end{tabular}
\caption{ \textbf{South Korea forecasting:} (a) Quarantine strength $Q(t)$ and (b) Effective reproduction number $R_{t}$, based on the neural network augmented SIR model. }\label{Koreaf}
\end{figure}

\begin{figure}
\centering
\begin{tabular}{cc}
\subfloat[]{\includegraphics[width=0.32\textwidth]{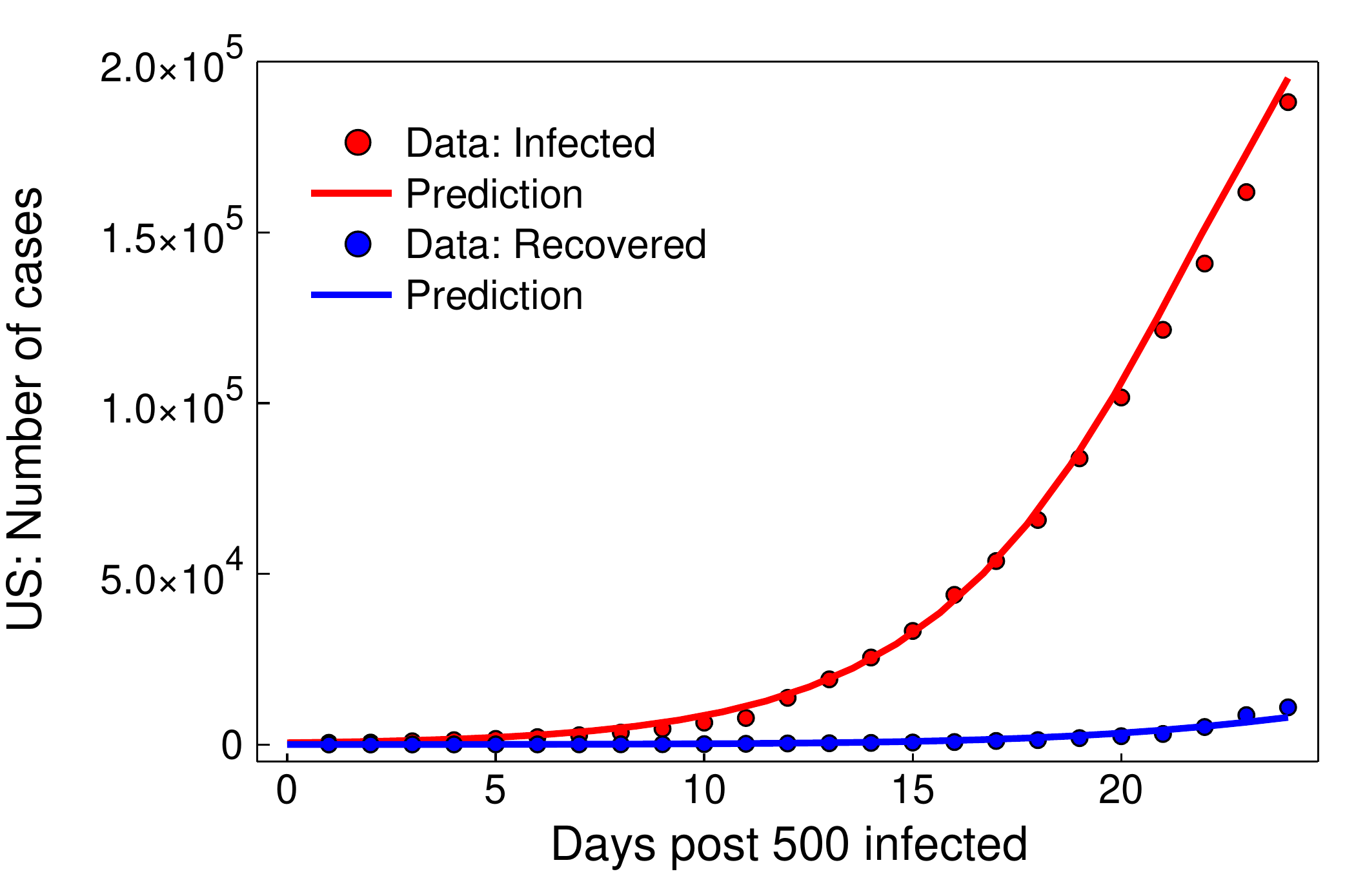}}
 \subfloat[]{\includegraphics[width=0.315\textwidth]{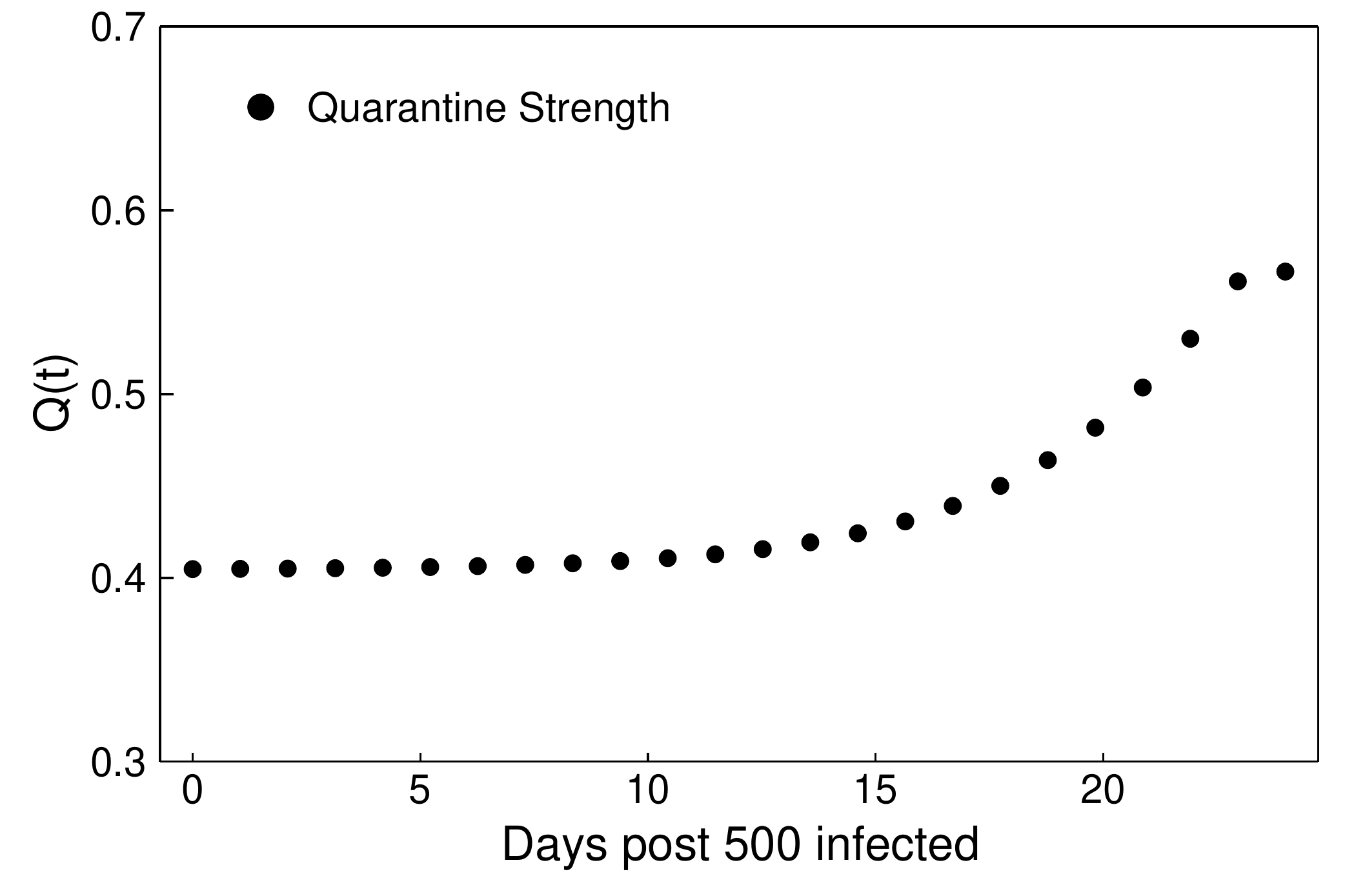}} 
  \subfloat[]{\includegraphics[width=0.315\textwidth]{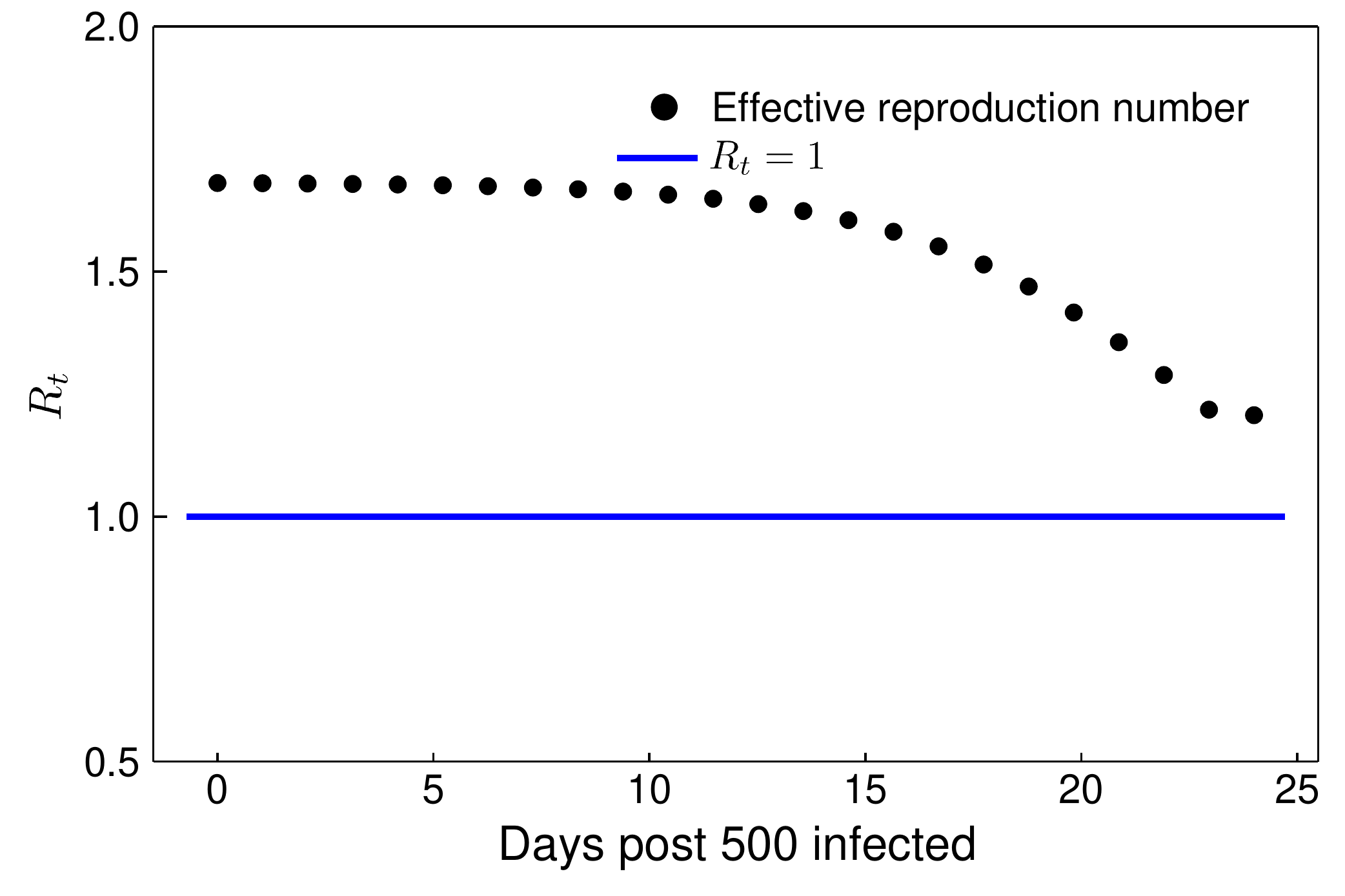}} 
\end{tabular}
\caption{\textbf{USA neural network model, day 0 = 8 March 2020:} (a) Estimation of the infected and recovered case count compared to the data post 8 March 2020 in USA. (b) Quarantine strength function $Q(t)$ learnt by the neural network. (c) Effective reproduction number $R_{t}$.}\label{USAn}
\end{figure}

\begin{figure}
\centering
\begin{tabular}{cc}
\subfloat[]{\includegraphics[width=0.4\textwidth]{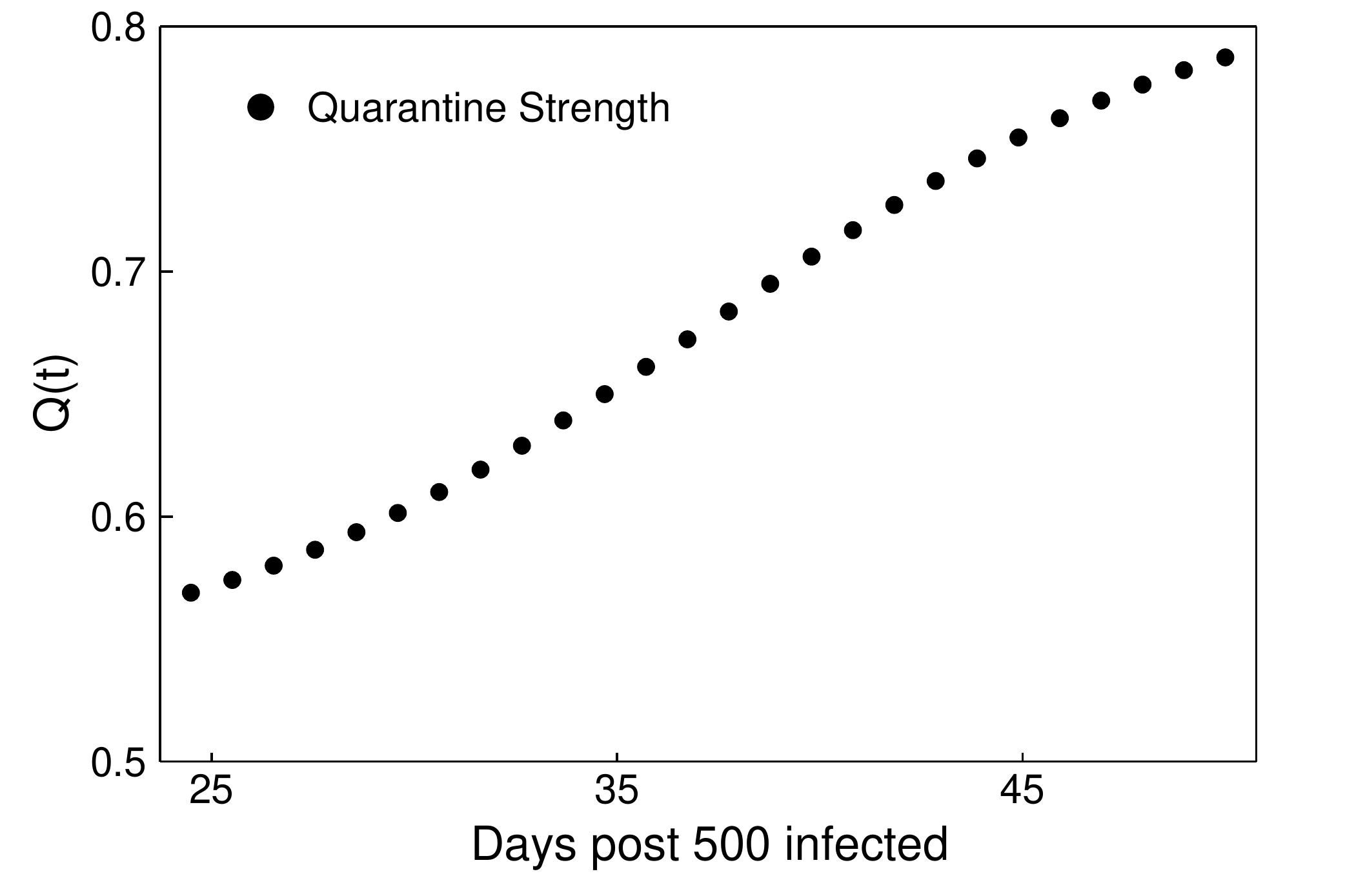}}
\subfloat[]{\includegraphics[width=0.42\textwidth]{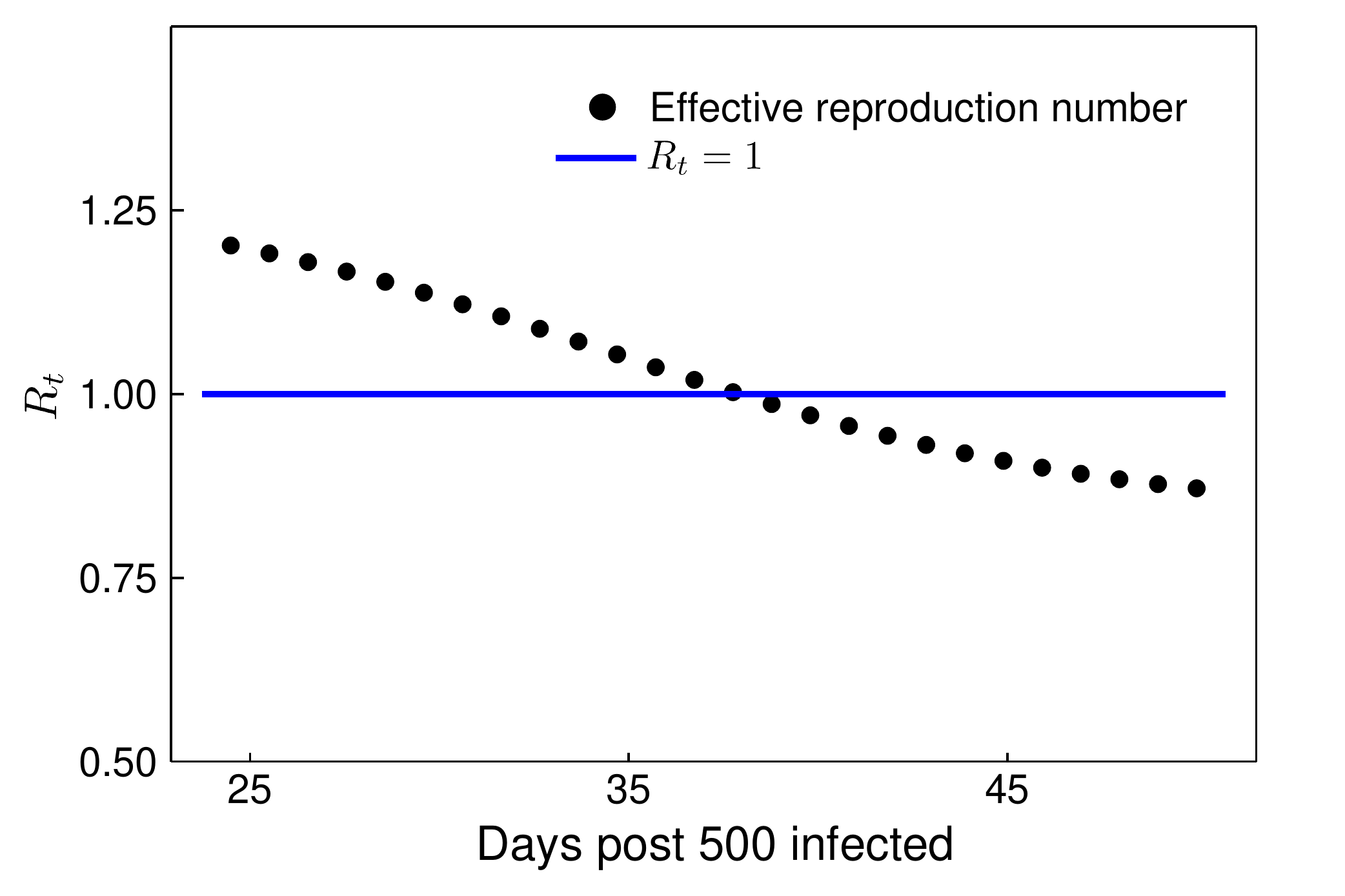}}
\end{tabular}
\caption{ \textbf{USA forecasting:} (a) Quarantine strength, $Q$ and (b) Effective reproduction number, $R_{t}$ in USA based on the neural network augmented SIR model. }\label{USAfn}
\end{figure}

\begin{figure}
\centering
\begin{tabular}{cc}
\subfloat[]{\includegraphics[width=0.42\textwidth]{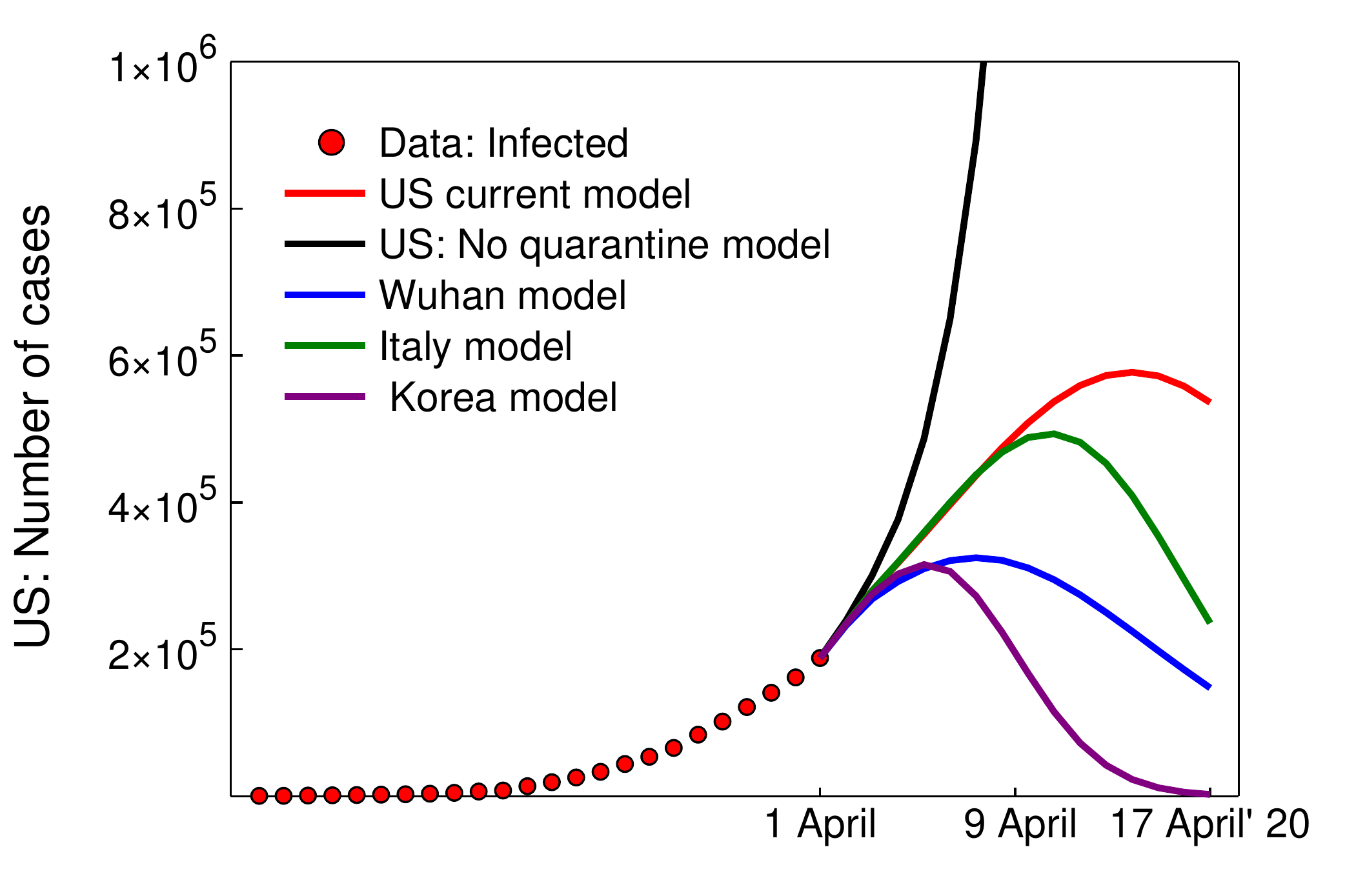}}
\subfloat[]{\includegraphics[width=0.41\textwidth]{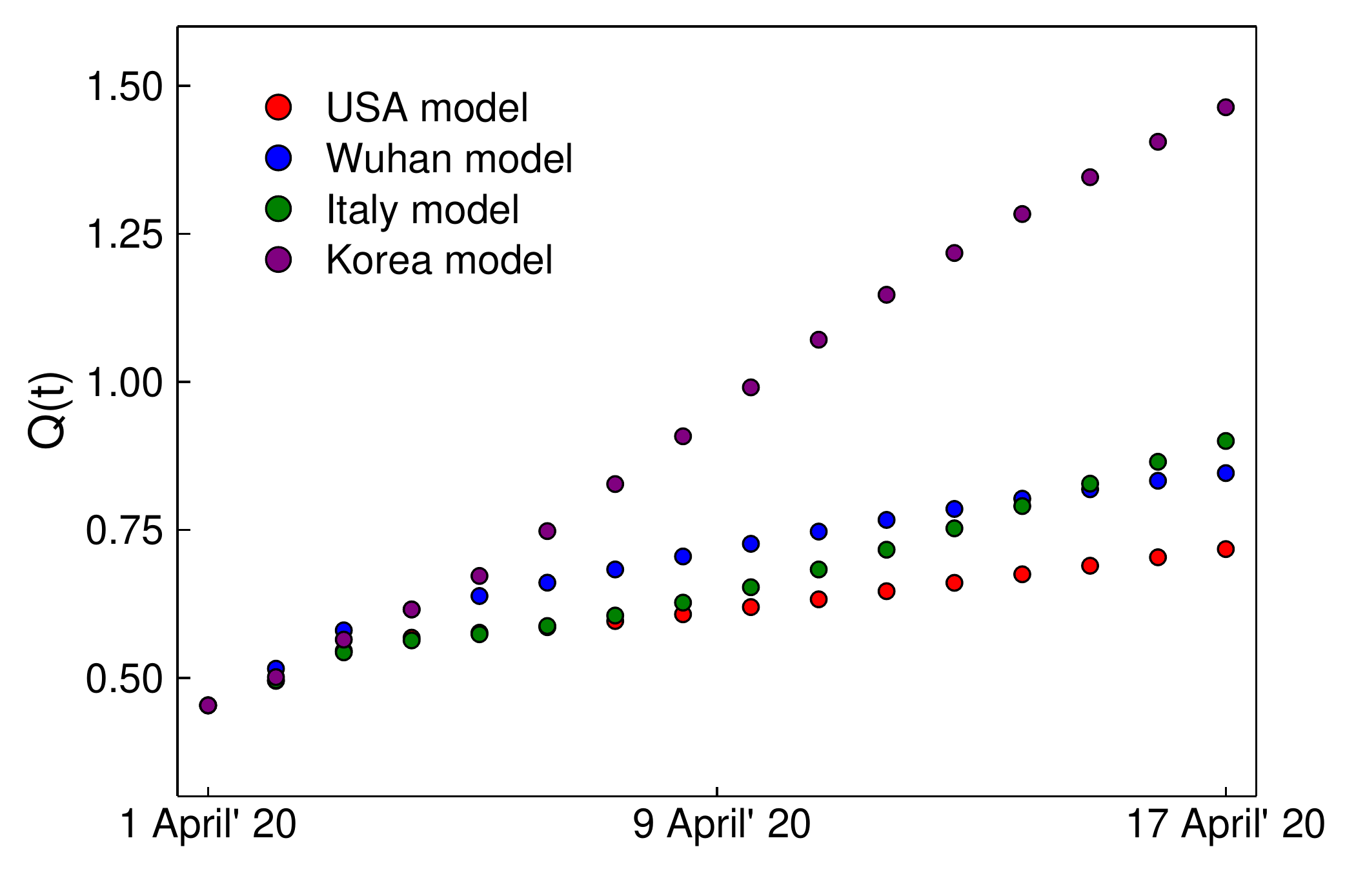}}

\end{tabular}
\caption{\textbf{USA mixed models forecasting, day 0 = 8 March 2020:} (a) Comparison of the forecast infected case count for USA after March $31^{\text{st}}$ according to the trained model for USA, or gradual adjustment to one of the quarantine control functions $Q(t)$ of Wuhan, Italy or Korea according to (\ref{Quar_USA}). (b) Adjusted quarantine strength function for USA according to (\ref{Quar_USA}).}\label{USAf}
\end{figure}

\section{Results}

Figure \ref{Wuhan-no-quar} shows results from the classical SEIR and SIR models applied to Wuhan data. Neither model can recover the stagnation seen in the actual infected number, about $30$ days post the detection of the $500^{\textrm{th}}$ infected case in Wuhan, {\it i.e.} $24^{\text{th}}$ January, 2020. The neural network model trained to include quarantine, on the other hand, does predict this stagnation; see below. 

Figures \ref{Wuhan-quar}-\ref{Koreaf} show results of our neural network models trained to include quarantine effects in the Wuhan, Italy, and South Korea regions, and respective predictions for the evolution of infections for approximately one month past the end of training. We use two parameters to quantify the results: the effective reproduction number $R_{t}$ and the effective quarantine strength $Q(t)$. Qualitatively, $Q(t)$ is inversely proportional to the average time it takes for an infected person to be quarantined and to further infect any susceptible individuals.

We trained the models using data starting from the dates when the $500^{\text{th}}$ infection was recorded in each region: $24^{\text{th}}$ January, $27^{\text{th}}$ February, and $22^{\text{nd}}$ February for Wuhan, Italy and South Korea, respectively; and up to about a month thereafter. The respective models, superimposed over the actual recorded data, are in Figures \ref{Wuhan-quar}a, \ref{Italy}a and \ref{Korea}a, generally showing good agreement. The respective forecasts are in Figures \ref{Wuhan-1month-forecast}, \ref{Italyf} and \ref{Koreaf}. As can be seen, the model with quarantine control included is able to capture the plateau in the infected case count, as opposed to the standard SIR and SEIR models shown in Figure \ref{Wuhan-no-quar}. In the case of Wuhan, because the onset was earliest, our forecast can be compared with recorded infection data in the period March $3^{\text{rd}}$ till March $24^{\text{th}}$, also showing good agreement with the actual observations of infection stagnation leading to $R_{t}<1$ during that period \citep{cyranoski2020china}.

In the case of the US, similarly to the other regions, {\it i.e.} starting when the $500^{\text{th}}$ infection was recorded on March $8^{\text{th}}$, we trained the model till the latest available data, {\it i.e} till $1$ April 2020. The infected case count estimated by our model shows a good match with the actual data (Figure \ref{USAn}a). Forecasting results for $Q(t), R_{t}$ for a period of $1$ month following the current US policy are in Figure \ref{USAfn}.
\iffalse
Under-reporting of the recovered case count in the USA \citep{NY}, leads to smaller recovery rate and hence a smaller fraction of the population being transferred from the infected compartment to the recovered compartment in the model described in (4.12) - (4.15). This leads to our model slightly over-estimating the infected case count (figure \ref{USAn}a). 
\fi
In Figure \ref{USAf}a we forecast the number of infections the US would experience starting from $1^{\textrm{st}}$ April if the US were to follow its current quarantine policy as opposed to gradually adjust to adopting the respective quarantine models learnt from the more reliable Wuhan, Italy and South Korea data. We arbitrarily set the adjustment period to $17$ days; i.e till $17$ April 2020. Figure \ref{USAf}b shows the effective quarantine rates $Q(t)$ obtained from (\ref{Quar_USA}) that would apply to the US during the respective adjustment period. Details are in the Materials \& Methods section.
\iffalse
We wish to emphasize that these data are rather speculative, as our model cannot really work for the US without reliable data on the recoveries $R(t)$ during the training period.
\fi

\section{Discussion}

The results show a generally strong correlation between strengthening of the quarantine controls, {\it i.e.} increasing $Q(t)$ as learnt by the neural network model; actions taken by the regions' respective governments; and decrease of the effective reproduction number $R_{t}$. For example, in Italy, government restrictions reportedly \citep{Italy_article} tightened during the week preceding mid March, which is also when our model shows a sharp increase in $Q(t)$ and corresponding decrease in $R_{t}$ (figures \ref{Italy}b, c). For Wuhan and South Korea, similar cusps in government interventions took place earlier, in the weeks leading to and after the end of January \citep{cyranoski2020china} and February, respectively \citep{Korea_article}. These cusps were also captured well by our model (Figures \ref{Wuhan-quar}b, c and \ref{Korea}b, c, respectively). Even for the USA, $Q(t)$ shows a stagnation till $20$ March 2020, after which it shows a sharp increase accompanied with a decrease in $R_{t}$ (Figures \ref{USAn}b, c), which is in alignment with the ramping up of government policies and quarantine interventions post mid March in the worst affected states like New York, New Jersey, California, and Michigan.

A comparative analysis of the quarantine strength function $Q(t)$ learned by the neural network for different countries reveals that Wuhan had the highest magnitude and South Korea had the highest growth rate of $Q(t)$. This can be attributed to the stringent government interventions and strict public health measures including immediate isolation and quarantine impositions in Wuhan and South Korea. This eventually resulted in the halting of infection spread and a corresponding $R_{t} <1 $ within a month for Wuhan (Figure \ref{Wuhan-quar}c) and within $20$ days for South Korea (figure \ref{Korea}c) after the first signs of a pandemic were recognized.

It is reported that the infected case count stagnated nation-wide in China by the beginning of March \citep{cyranoski2020china} and in South Korea by the end of March \citep{Korea_flattening}; which eventually led to a stagnation in the quarantine interventions employed in these countries. This is in general qualitative agreement with our forecasting results which show a plateau in $Q(t)$ and $R_{t}$  at $R_{t} < 1$: Figures \ref{Wuhan-1month-forecast}a,b and \ref{Koreaf}a,b. In Italy, as of March $20^{\text{th}}$, $I(t)$ is appearing to be linear (Figure \ref{Italy}a), which is consistent with lower rates of infections being actually reported \citep{Italy_flattening} and can be taken as a precursor to stagnation. It is also consistent with adoption of strict movement restrictions by the government shortly before the March $20^{\text{th}}$ date. We forecast that, for Italy, $R_{t}$ will drop below $1$ and $Q(t), R_{t}$ both will stagnate between mid to end of April 2020 (Figure \ref{Italyf}a, b) indicating halting of the spread of infection.

Owing to the relaxed quarantine and isolation policies in the US in its initial stages post the infection spread, our model converges to $Q(t) \approx 0.4 - 0.6$ (Figure \ref{USAn}b) which is the smallest compared to other regions. Even though the effective $R_{t}$ is still greater than $1$ as of April $1$ 2020 (Figure \ref{USAn}c), its growth has started to show a decreasing trend and we expect the infection to start showing stagnation with $R_{t}<1$ by $20$ April 2020 if the current US policies continue without change (Figures \ref{USAfn}b, \ref{USAf}a). This will be accompanied by a continuous ramp up of quarantine policies (Figure \ref{USAfn}a). At its peak, we forecast the infected count to reach approximately 600,000 before stagnation, again assuming no change in US policies. Our mixed model analysis for USA, employing $Q(t)$ learnt from the models of Wuhan, Italy and South Korea in the USA model starting from $1$ April 2020, reveals that stronger quarantine policies (Figure \ref{USAf}b) might lead to an accelerated plateauing in the infected case count, as shown in Figure \ref{USAf}a, and subsequently smaller infected case count. On the other hand, in agreement with National Institute of Allergy and Infectious Diseases estimates \citep{Fauci}, we forecast that relaxing or abandoning the quarantine policies gradually over the period of the next $17$ days may well lead to $\sim 1$ million infections without any stagnation in the infected case count (Figure \ref{USAf}a) by mid April 2020.

\section{Materials and Methods}

\textbf{Model 1: Without quarantine control} 

The classic SEIR epidemiological model has been employed in a number of prior studies, such as the SARS outbreak \cite{SEIR1, SEIR2, SEIR3} as well as the Covid outbreak \cite{read2020novel, tang2020estimation, wu2020nowcasting}. The entire population is divided into four sub-populations:  susceptible $S$;  exposed $E$;  infected $I$; and  recovered $R$. The sub-populations' evolution is governed by the following system of four coupled nonlinear ordinary differential equations \citep{SEIR3, wang2020phase}

\begin{align}
    \ddtt{S} & = -\frac{\beta \: S(t) \: I(t)}{N} \\
    \rule{0cm}{4ex} \ddtt{E} & = \frac{\beta \: S(t) \: I(t)}{N} - \sigma E(t) \\
    \rule{0cm}{4ex} \ddtt{I} & = \sigma E(t) - \gamma I(t) \\
    \rule{0cm}{4ex} \ddtt{R} & = \gamma I(t).
\end{align}
Here, $\beta$, $\sigma$ and $\gamma$ are the exposure, infection and recovery rates, respectively, and are assumed to be constant in time. The total population $N=S(t)+E(t)+I(t)+R(t)$ is seen to remain constant as well; that is, births and deaths are neglected. The recovered population is to be interpreted as those who can no longer infect others; so it also includes  individuals deceased due to the infection. The possibility of recovered individuals to become reinfected is accounted for by SEIS models \citep{mukhopadhyay2008analysis}, but we do not use this model here, as the reinfection rate for Covid-19 survivors is considered to be negligible as of now. 

The simpler SIR model neglects exposure, assuming instead direct transition from susceptible to infected; it is described by three coupled nonlinear ordinary differential equations as 
\begin{align}
    \ddtt{S} & = -\frac{\beta \: S(t) \: I(t)}{N} \\
    \rule{0cm}{4ex} \ddtt{I} & = \frac{\beta \: S(t) \: I(t)}{N} - \gamma I(t) \\
    \rule{0cm}{4ex} \ddtt{R} & = \gamma I(t).
\end{align}
Here, $\beta$ is the infection rate. 

The reproduction number $R_{t}$ in the SEIR and SIR models is defined as
\begin{equation}
    R_{t} = \cfrac{\beta}{\gamma}.
\end{equation}
An important assumption of the SEIR and SIR  models is homogeneous mixing among the subpopulations. Therefore, they cannot account for social distancing or mass quarantine effects. Additional assumptions are uniform susceptibility and disease progress for every individual; and that no spreading occurs through animals or other non-human means. Alternatively, the  models may be interpreted as quantifying the statistical expectations on the respective mean populations, while deviations from the model's assumptions contribute to statistical fluctuations around the means. 

We applied both SEIR and SIR models to the case of Wuhan only. The results are shown in Figure \ref{Wuhan-no-quar} and verify that these models fail to predict the early arrest of infectious spread due to quarantine policies. 

\noindent {\bf Initial conditions}\ The starting point $t  =0$ was the day at which $500$ infected cases were detected, {\it i.e.}, $I(t=0)=500$. The initial number of susceptible individuals was $ S(t=0) = 11$ million, Wuhan's population. The initial exposed population was assumed to be $E(t=0) = 20 \times I(t=0)$ in accordance with \citep{read2020novel, wang2020phase} and the number of recovered individuals was set to a very small value $R(t=0) \approx 10$. 

\noindent {\bf Parameter estimation}\ The time resolved data for the infected, $I_{\textrm{data}}$ and recovered, $R_{\textrm{data}}$  case count for Covid-19 was obtained from Chinese National Health Commission. The optimal values of the parameters $\beta, \gamma$ in the SIR model and $\beta, \sigma, \gamma$ in the SEIR model were obtained by performing a local adjoint sensitivity analysis \citep{cao2003adjoint, Rack2} of the ODE problems in these models, by minimizing the mean square error loss function $L (\beta, \sigma, \gamma)$ defined as
\begin{equation}\label{loss}
    L (\beta, \sigma, \gamma) = ||\textrm{log}(I(t)) - \textrm{log}(I_{\textrm{data}}(t))||^{2} + ||\textrm{log}(R(t)) - \textrm{log}(R_{\textrm{data}}(t))||^{2} 
\end{equation}
The minimization procedure was performed using the ADAM optimizer \citep{kingma2014adam} for $500 \sim 1000$ iterations. The optimum values were used to produce figure \ref{Wuhan-no-quar}.

\textbf{Model 2: With quarantine control} 

To study the effect of quarantine control globally, we start with the SIR epidemiological model. This choice is to minimize the number of free parameters in the model and avoid overfitting. To include quarantine control in the modelling, we augment the SIR model by introducing a time varying quarantine strength term $Q(t)$ and a quarantined population $T(t)$  
\[
T(t) = Q(t) \times I(t).
\]
Thus, the new effective reproduction rate becomes
\begin{equation}
    R_{t} = \cfrac{\beta}{\gamma + Q(t)}.
\end{equation}
Since $Q(t)$ does not follow from first principles and is highly dependent on local quarantine policies, we devised a neural network-based approach to approximate it. 

Recently, it has been shown that neural networks can be used as function approximators to recover unknown constitutive relationships in a system of coupled ordinary differential equations \citep{Rackauckas20, Rack2}. Following this principle, we represent $Q(t)$ as a $n$ layer-deep neural network with weights $W_{1}, W_{2} \ldots W_{n}$, activation function $r$ and the input vector $U = (S(t), I(t), R(t), T(t))$ as
\begin{equation}\label{NN-1}
   Q(t) = r\left( W_{n} r\left( W_{n-1} \ldots r\left( W_{1} U\right)\right)\right) \equiv \text{NN}(W, U)
\end{equation}

For the implementation, we choose a $n=2$-layer densely connected neural network with $10$ units in the hidden layer and the $\textrm{ReLU}$ activation function. This choice was because we found sigmoidal activation functions to stagnate. The final model was described by $63$ tunable parameters. The neural network architecture schematic is shown in the attached Supplementary Information. The governing coupled ordinary differential equations for the augmented SIR model are 
\begin{align}\label{model_augment}
    \ddtt{S} & = -\frac{\beta \: S(t) \: I(t)}{N} \\
    \rule{0cm}{4ex} \ddtt{I} & = \frac{\beta \: S(t) \: I(t)}{N} - \left(\rule[-0.5ex]{0cm}{2.5ex} \gamma+Q(t)\right) I(t) =
    \nonumber \\
    & = \frac{\beta \: S(t) \: I(t)}{N} - \left(\rule[-0.5ex]{0cm}{2.5ex} \gamma+\text{NN}(W, U)\right) I(t) \\
    \rule{0cm}{4ex} \ddtt{R} & = \gamma I(t) \\
    \rule{0cm}{4ex} \ddtt{T} & = Q(t)\: I(t) = \text{NN}(W, U)\: I(t).
\end{align}
The quarantined population is initialized to a very small value $T(t=0)\approx 10$ and, thereafter, the neural network learns how to approximate it based on the local data from each region under study.

\noindent{\bf Initial conditions}
The starting point $t  =0$ for each simulation was the day at which $500$ infected cases were detected, {\it i.e.} $I_{0} = 500$. The number of susceptible individuals was assumed to be equal to the respective regional populations, {\it i.e.} $S(t=0) = 11$ million, $60$ million, $52$ million and $327$ million for Wuhan, Italy, South Korea and USA respectively. Also, in all simulations, the number of recovered individuals was initialized to a small number $R(t=0) \approx 10$.  

\noindent{\bf Parameter estimation}
The time resolved data for the infected, $I_{\textrm{data}}$ and recovered, $R_{\textrm{data}}$ for each locale considered is obtained from the Center for Systems Science and Engineering (CSSE) at Johns Hopkins University. The neural network-augmented SIR ODE system was trained by minimizing the mean square error loss function 
\begin{equation}\label{loss2}
    L_{\text{NN}} (W, \beta, \gamma) = ||\textrm{log}(I(t)) - \textrm{log}(I_{\textrm{data}}(t))||^{2} + ||\textrm{log}(R(t)) - \textrm{log}(R_{\textrm{data}}(t))||^{2} 
\end{equation}
that includes the neural network's weights $W$. Minimization was carried out through local adjoint sensitivity analysis \citep{cao2003adjoint, Rack2} following a similar procedure outlined in \citet{Rackauckas20} and implemented using the ADAM optimizer \citep{kingma2014adam} for  $300 \sim 500$ iterations. To avoid over-fitting, the training is stopped when the loss function, $L$ is seen to stagnate and the first derivative $I'(t), R'(t)$ is seen to match that of the data. The training loss curves for all regions are shown in the attached Supplementary Information. The plots in figures \ref{Wuhan-quar}, \ref{Italy}, \ref{Korea} and \ref{USAn} were obtained by following this approach separately for each locale, training the respective neural networks according to data available for the periods January $24^{\text{th}}$ - March $3^{\text{rd}}$  for Wuhan, February $24^{\text{th}}$- March $23^{\text{rd}}$   for Italy, February $22^{\text{nd}}$-March $17^{\text{th}}$  for Korea, and March $8^{\text{th}}$- April $1^{\text{st}}$  for the US. It should be noted that under-reporting of the recovered case count, $R(t)$ in the USA \citep{NY}, leads to smaller recovery rate and hence a smaller fraction of the population being transferred from the infected compartment to the recovered compartment in the model described in (4.12) - (4.15). Thus, simultaneous training of $W, \beta, \gamma$ as described above leads to our model over-estimating the infected case count. As a result, for the USA, we first find the optimal $\gamma$ by minimizing (\ref{loss2}), and then use this value of $\gamma$ for minimizing the loss function, $L(W, \beta) = ||\textrm{log}(I(t))- \textrm{log}(I_{\textrm{data}}(t))||^{2}$. Such an independent optimization procedure for estimating $\gamma$ may lead to small errors in the estimation of $R(t)$. Such errors are seen to be negligible in this case (figure \ref{USAn}a), thus validating this approach for the USA.

The estimates of $\beta$, $\gamma, Q(t)$ obtained using this procedure are shown in table \ref{tab:betagamma}. Table also shows the intervention efficiency defined as the number of days elapsed between detection of $500^{\textrm{th}}$ case and the first time when the effective reproduction number reached $R_{t}<1$ in the chosen locale.

Forecasts in figures \ref{Wuhan-1month-forecast}, \ref{Italyf}, \ref{Koreaf} were obtained by using the sub-population data on the final days of their respective training periods to initialize the {\em trained} neural network models for Wuhan, Italy and Korea. For figure \ref{USAf}, the forecasts were obtained by similarly initializing the model but subsequently in the post April 1$^{\text{st}}$ period adjusting the quarantine model gradually over $17$ days till $10$ April according to
\begin{equation}\label{Quar_USA}
Q(t) = \frac{17-t}{17}\:Q_{\text{USA}}(t) + \frac{t}{17}\:Q_j(t), \qquad (t=0\text{~on April~}1^{\text{st}})
\end{equation}
where $j=$ Wuhan, Italy, Korea for the respective assumed quarantine policy adoptions.

\begin{table}
\begin{center}
\def~{\hphantom{0}}
  \begin{tabular}{|M{2.5cm}|M{1.25cm}|M{1.25cm}|M{2.5cm}|M{2.5cm}|} \hline
 Region		 & $\beta$ 	& $\gamma$  & Range of $Q(t)$ & Intervention efficiency	 \\ \hline
  Wuhan	& $1$ 	& $0.023$ & $0.8 - 1.1$ & $30$\\
  Italy	& $0.74$ 	& $0.032$ & $0.4-0.7$ & $27$\\	
  South Korea	& $0.68$ 	& $0.004$ & $0.4 - 0.8$  & $20$\\	
  US	& $0.69$ 	& $0.008$ & $0.4 - 0.6$ & \ \ \ \ \  37 \newline (Forecasted)\\	\hline
  \end{tabular}
  \caption{Table shows infection and recovery rates $\beta$ and $\gamma$, respectively, and the range of  quarantine strength function, $Q(t)$ obtained from minimising (\ref{loss2}); along with the intervention efficiency defined as the number of days elapsed between detection of $500^{\textrm{th}}$ case and the first time when the effective reproduction number reached $R_{t}<1$.}
    \label{tab:betagamma}
  \end{center}
\end{table}
\vspace{1ex}

\begin{figure}[H]
\centering
\begin{tabular}{cc}
\includegraphics[width=0.4\textwidth]{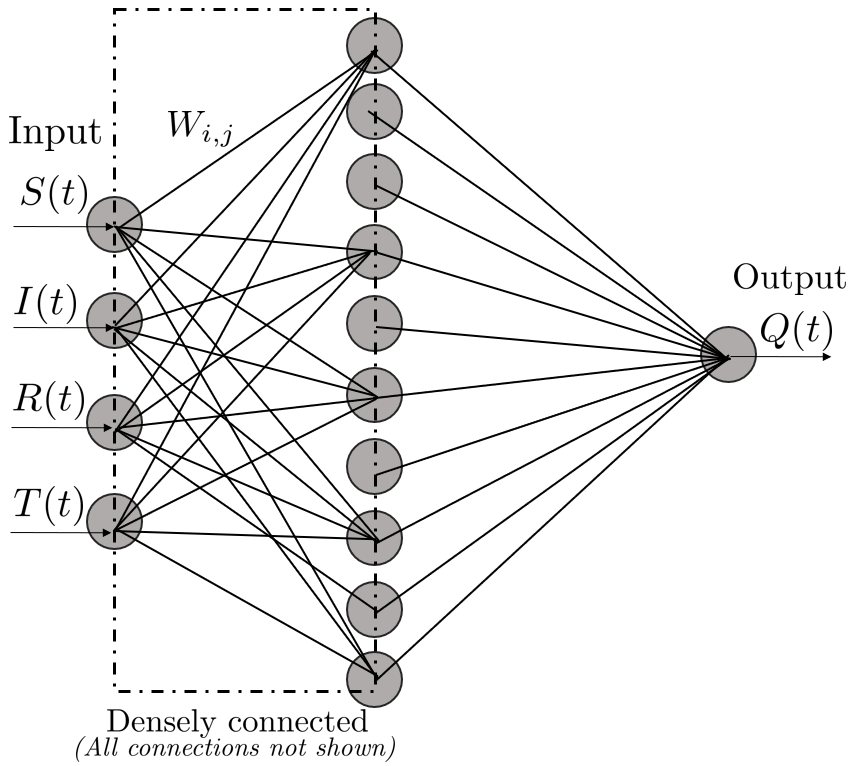}
\includegraphics[width=0.4\textwidth]{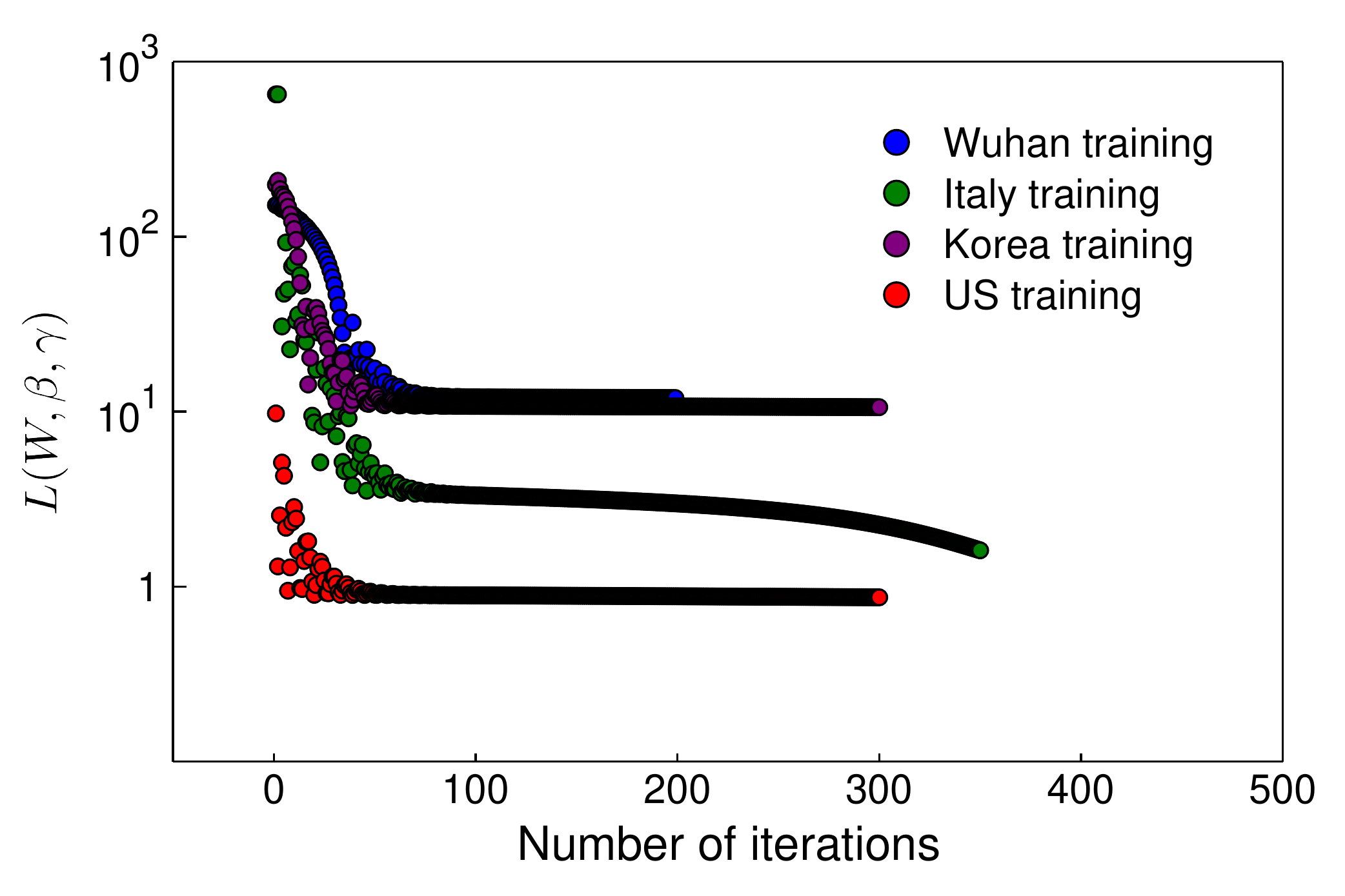}
\end{tabular}
\caption{ (a) Neural network architecture used: $2$ layers with $10$ units in the hidden layer. (b) Training loss, $L(W, \beta, \gamma)$ according to (4.16) for all regions considered.}\label{supp1}
\end{figure}

\noindent {\bf Conflicts of Interest}
The authors declare no conflicts of interest. \newline\newline
\noindent {\bf Data collection} 
Data for the infected and recovered case count in Wuhan is obtained from the data released by the Chinese National Health Commission. Infected and recovered count data for Italy, South Korea and USA is obtained from the Center for Systems Science and Engineering (CSSE) at Johns Hopkins University. \newline\newline
\noindent {\bf Acknowledgment}
This effort was partially funded by the Intelligence Advanced Reseach Projects Activity (IARPA.) We are grateful to Haluk Akay, Hyungseok Kim and Wujie Wang for helpful discussions and suggestions. \newline\newline
\noindent {\bf Supplementary Notes: } 
Supplementary figures are shown in figure \ref{supp1}.
\bibliography{Paper_Draft1}
\bibliographystyle{jfm}
\end{document}